\documentclass[sigconf]{acmart}
\usepackage[normalem]{ulem}
\usepackage{latexsym}
\usepackage{microtype}

\usepackage{amsmath}
\usepackage{graphicx}

\usepackage{subfigure}
\usepackage{stfloats}

\usepackage{makecell}
\usepackage{multicol}
\usepackage{multirow}

\AtBeginDocument{%
  \providecommand\BibTeX{{%
    \normalfont B\kern-0.5em{\scshape i\kern-0.25em b}\kern-0.8em\TeX}}}

\copyrightyear{2023}
\acmYear{2023}
\setcopyright{acmlicensed}\acmConference[CIKM '23]{Proceedings of the 32nd
ACM International Conference on Information and Knowledge
Management}{October 21--25, 2023}{Birmingham, United Kingdom}
\acmBooktitle{Proceedings of the 32nd ACM International Conference on Information and Knowledge Management (CIKM '23), October 21--25, 2023, Birmingham, United Kingdom}
\acmPrice{15.00}
\begin{document}

\title{Text Matching Improves Sequential Recommendation by Reducing Popularity Biases}


\author{Zhenghao Liu}
\authornote{indicates equal contribution.}
\authornote{indicates corresponding author.}
\affiliation{
  \institution{Northeastern University}
  \city{Shenyang}
  \country{China}}
\email{liuzhenghao@mail.neu.edu.cn}

\author{Sen Mei}
\authornotemark[1]
\affiliation{
  \institution{Northeastern University}
  \city{Shenyang}
  \country{China}}
\email{meisen@stumail.neu.edu.cn}

\author{Chenyan Xiong}
\affiliation{
  \institution{Carnegie Mellon University}  \city{Pittsburgh}
  \country{United States}}
  \email{cx@cs.cmu.edu}

\author{Xiaohua Li}
\affiliation{
  \institution{Northeastern University}
  \city{Shenyang}
  \country{China}}
  \email{lixiaohua@mail.neu.edu.cn}

\author{Shi Yu}
\affiliation{%
  \institution{Tsinghua University}
  \city{Beijing}
  \country{China}}
  \email{yus21@mails.tsinghua.edu.cn}

\author{Zhiyuan Liu}
\affiliation{%
  \institution{Tsinghua University}
  \city{Beijing}
  \country{China}}
  \email{liuzy@tsinghua.edu.cn}

\author{Yu Gu}
\affiliation{
  \institution{Northeastern University}
  \city{Shenyang}
  \country{China}}
   \email{guyu@mail.neu.edu.cn}

\author{Ge Yu}
\affiliation{
  \institution{Northeastern University}
  \city{Shenyang}
  \country{China}}
    \email{yuge@mail.neu.edu.cn}



\begin{abstract}
This paper proposes \textbf{T}ext m\textbf{A}tching based \textbf{S}equen\textbf{T}ial r\textbf{E}commenda-tion model (TASTE), which maps items and users in an embedding space and recommends items by matching their text representations. TASTE verbalizes items and user-item interactions using identifiers and attributes of items. To better characterize user behaviors, TASTE additionally proposes an attention sparsity method, which enables TASTE to model longer user-item interactions by reducing the self-attention computations during encoding. Our experiments show that TASTE outperforms the state-of-the-art methods on widely used sequential recommendation datasets. TASTE alleviates the cold start problem by representing long-tail items using full-text modeling and bringing the benefits of pretrained language models to recommendation systems. Our further analyses illustrate that TASTE significantly improves the recommendation accuracy by reducing the popularity bias of previous item id based recommendation models and returning more appropriate and text-relevant items to satisfy users. All codes are available at \url{https://github.com/OpenMatch/TASTE}.

%Dynamic modeling of user-item interaction history is an important and challenging task in the field of recommendation system. In order to improve the prediction ability of the model, the existing approach adopts the method of integrating side information, but there are still some problems. Most of the existing models are based on id embedding, This requires rich data support to learn enough good embedding, so there is naturally a cold start problem. In order to fuse side information, various complex models are designed, and the fusion method sometimes brings additional noise, resulting in poor fusion effect. Due to the strong encoding ability of the pretrained language model, we have reason to believe that we can learn a better item representation by using the description information of the items, and further extend it. We have adopted the full text method to model the sequence and item representation, and transformed the recommendation task into a dense retrieval task. Furthermore, we explored the long text modeling, and we used FID to model the long text. Experiments on two real world datasets show that the proposed method is superior to the existing state-of-the-art models.All source codes of the work are available at xxxx.
\end{abstract}






\maketitle

\section{Introduction}
Sequential recommendation systems~\cite{chen2018sequential,Sun2019Bert4rec,Zhou2020s3} dynamically recommend the next items for users, which play a crucial role in lots of web applications, such as Yelp, TikTok, Amazon, etc. These recommendation systems model user behaviors by employing different neural architectures to learn the dependency among items in the user-item interactions~\cite{Jiaxi2018Personalized,Hidasi2015session,Chang2021graph,Wang2018self}. They usually represent items using ids, randomly initialize item id embeddings, and optimize these embeddings using the signals from user-item interactions. 

\begin{figure}[t]
    \centering 
    \subfigure[The Item Distribution with User-Interacted Frequency.] { 
    \label{fig:longtail:freq} 
     \includegraphics[width=0.48\linewidth]{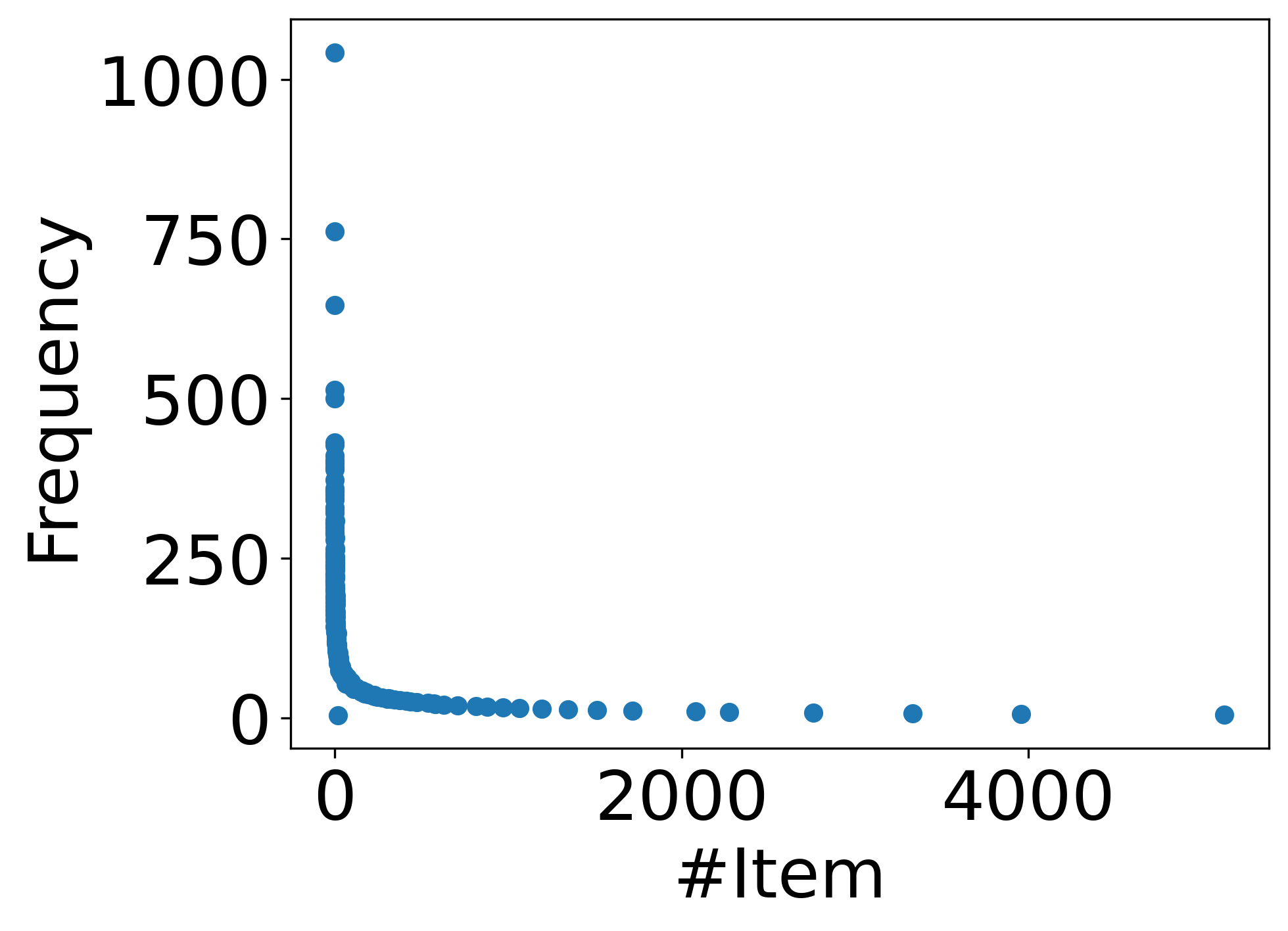}}
     \subfigure[Item Statistics.] { 
     \label{fig:longtail:num} 
    \includegraphics[width=0.48\linewidth]{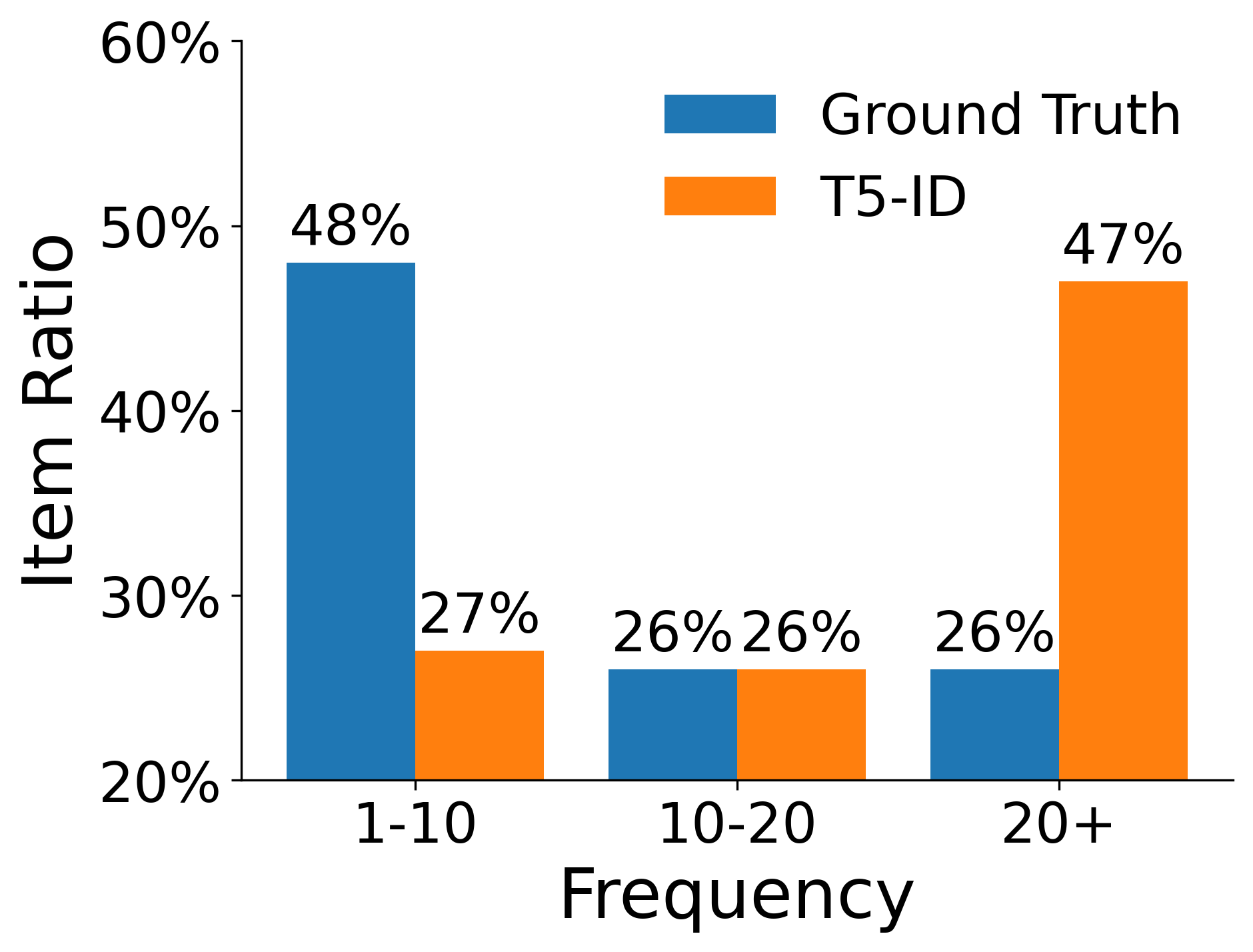}}
    \caption{Item Distributions with Different User-Item Interaction Frequencies. We sort items according to the user-item interaction frequencies and plot the user-interacted frequency distribution of items in Figure~\ref{fig:longtail:freq}. Then we conduct item statistics of different user-item interaction frequencies. The ground truth and top-5 items that are predicted by T5-ID are shown in Figure~\ref{fig:longtail:num}. T5-ID is an item id based recommendation model, which predicts the id of the next item. The items are divided into different groups according to the frequency.}
    \label{fig:longtail}
    %\vspace{-1em}
\end{figure}

% \input{figure/template.tex}

% As shown in Figure~\ref{fig:popbias}, given a user-item interaction sequence ``1914 $\rightarrow$ $\cdots$ $\rightarrow$ 6523 $\rightarrow$ 9084 '', we aim to predict the ``11322'' item, which is related to the skin care. However, these item id embedding based sequential recommendation systems usually tend to recommend popular items that are more frequently interacted with users but are less relevant to previously visited items. 

\begin{figure}[t]
    \centering 
    \subfigure[Item ID based Model (T5-ID).] { 
    \label{fig:popbias:id} 
     \includegraphics[width=0.48\linewidth]{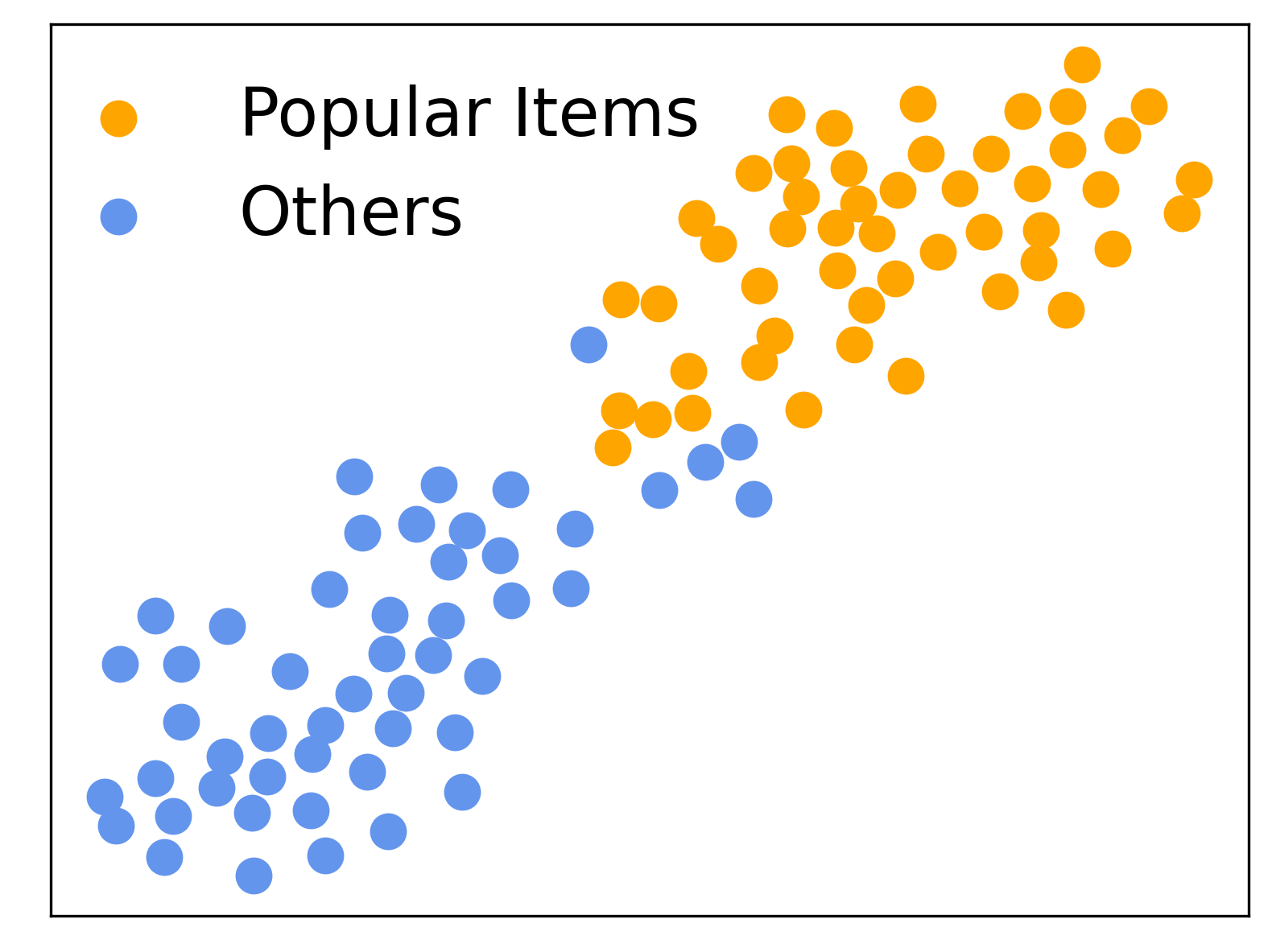}}
     \subfigure[Text Matching based Model (TASTE).] { 
     \label{fig:popbias:text} 
    \includegraphics[width=0.48\linewidth]{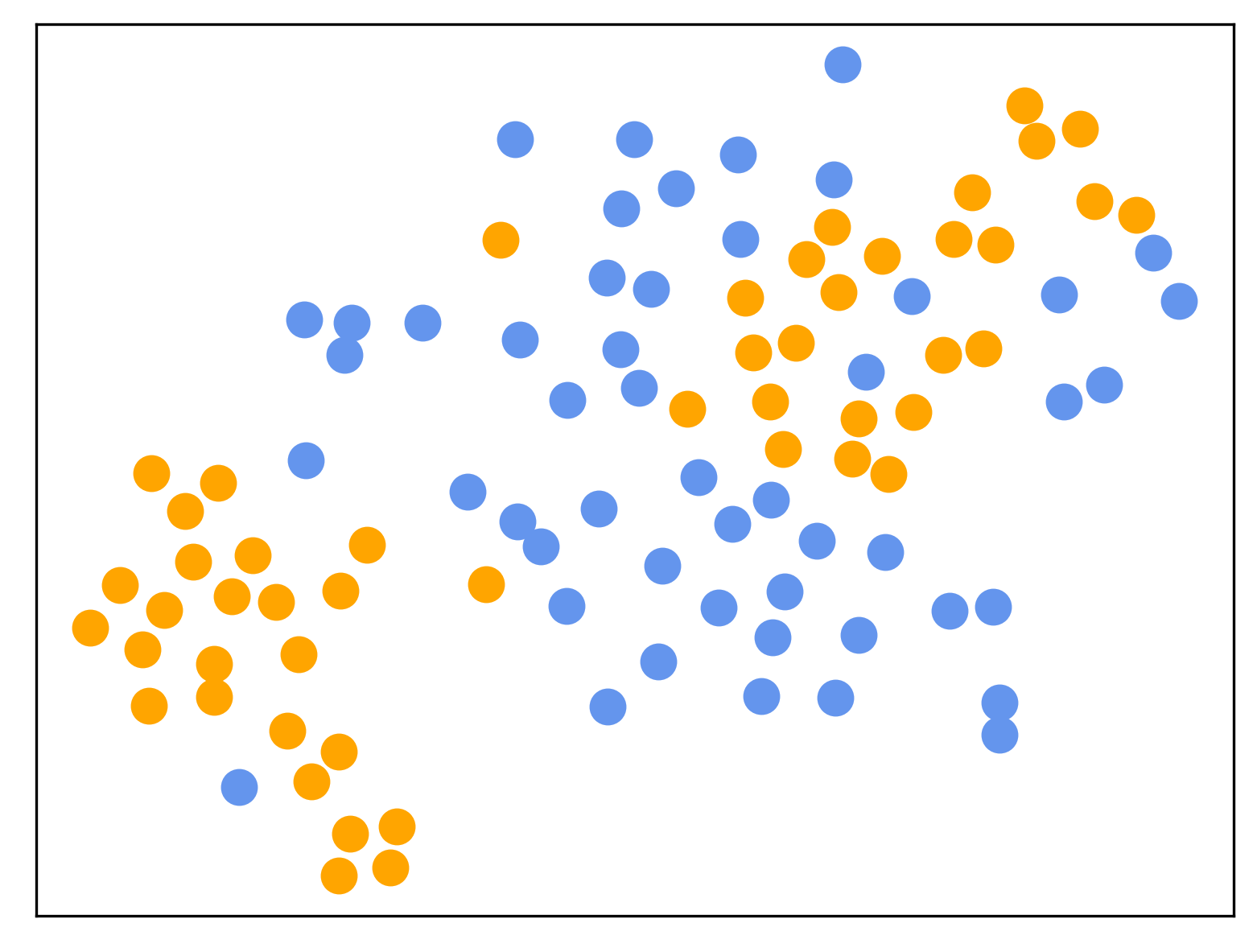}}
    \caption{Embedding Visualization of Items in Beauty. We randomly select 50 items from the popular item set and select 50 items from the rest. We use t-SNE to visualize the item embedding spaces of the item id based recommendation model and text matching based recommendation model in Figure~\ref{fig:popbias:id} and Figure~\ref{fig:popbias:text}, respectively.}
    \label{fig:popbias}

\end{figure}

% Existing recommendation systems usually represent items using ids, randomly initialize item id embeddings, and optimize these embeddings using the signals from user-item interactions. 
Existing item id based recommendation systems usually face the popularity bias problem~\cite{abdollahpouri2020multi,abdollahpouri2019unfairness,zhu2021popularity}. As shown in Figure~\ref{fig:longtail:freq}, the distribution of items in recommendation benchmarks is long-tailed, making the item id based recommendation model (T5-ID) usually face the ``Cold Start'' problem~\cite{lam2008addressing,schein2002methods}. In Figure~\ref{fig:longtail:num}, about 74\% golden items are interacted less than 20 times with users, but T5-ID returns more popular items in the recommendation results. It results in the popularity bias--\textit{just follow the crowd and return popular items as results}. Furthermore, we visualize the embedding distribution of T5-ID in Figure~\ref{fig:popbias:id}. T5-ID learns a non-smooth anisotropic embedding space~\cite{li2020sentence,gao2019representation,qiu2022contrastive}, which makes the popular items and other items distinguished. 
As shown in Figure~\ref{fig:popbias:text}, TASTE which represents items with full texts provides some opportunities to alleviate the popularity bias by mixing up the embeddings of popular items and others. It helps the recommendation system return more text-relevant but long-tailed items for users by matching text representations of users and items. %Long-term user-item interaction indeed helps to characterize the user behaviors more accurately~\cite{pi2019practice}, but it challenges the text matching based recommendation models to encode the long text representations of verbalized user-item interactions. %In this case, we represent both items and the user-item interaction history using text sequences of item identifiers and item attributes.%, such as names; prices, categories, and brands of products; addresses and categories of locations item attributes.

%The mainstream recommendation system models are based on the item id as the identifier. Due to the problems of long tail items and few user interaction records in the recommendation system~\cite{Yuan2020com}, leading to the cold start problem of the recommendation system. The id-based model maps item to id, resulting in the loss of the content information of the item itself. So some work is now trying to replace the id identifier to solve these problems~\cite{}. Item has text description information, such as . So these works try to use item text information as identifier. Pre-trained language models such as Bert~\cite{devlin2018bert}, T5~\cite{raffel2020exploring}, have demonstrated their powerful text encoding capabilities in the field of natural language processing.

In this paper, we propose \textbf{T}ext m\textbf{A}tching based \textbf{S}equen\textbf{T}ial r\textbf{E}commendation (TASTE), which represents items and users with texts, establishes relevance between them by matching their text representations and alleviates the popularity bias of previous item id based recommendation models. TASTE verbalizes users and items by designing some prompts~\cite{liu2023pre} and filling up the templates with item ids and item attributes. The item information provides text-matching clues to model the dependency and relevance among users and items. TASTE also proposes an attention sparsity encoding technology to break the max length boundary of language models~\cite{beltagy2020longformer} and encode long text representations of user-item interactions. It separates the user-item interaction into different sessions, independently encodes the text representation of each session, and reduces the attention computation. 
%It separate the text sequence of user interaction history as subsequences, regards these subsequences as interaction sessions, and reduce the self-attention computations between these subsequences. 

Our experiments demonstrate that TASTE surpasses previous sequential recommendation baselines~\cite{Xie2022DIF-SR,Sun2019Bert4rec,Zhou2020s3} with over 18\% improvements on Yelp and Amazon Product datasets. Our analyses show that, compared with item id based recommendation models, TASTE shows its ability in alleviating the popularity bias problem and recommends more than 18\% long-tail items for users compared to T5-ID. Our text based recommendation modeling enables TASTE to model the relevance between users and items by capturing the text matching signals. It helps to learn more effective representations for these long-tail items, alleviates the ``cold start'' problem, and makes TASTE return more text-relevant items as the recommendation result. Thrives on our attention sparsity method, TASTE has the ability to reduce the GPU memory and achieves more than 2\% improvements by modeling longer user-item interaction history. All experimental results show that the behavior modeling of user purchases and visits starts to be benefited from pretrained language models.

% Sequential recommendation system also has another challenge, long sequence encoding. With the rapid development of the Internet, users have accumulated a large number of behaviors on the Internet. Taking Taobao as an example, they reported that 23\% of users had more than 1000 behaviors in Taobao apps within six months.It is very important to capture user preferences by making better use of more user behavior sequence information. At the same time, due to the adoption of text encoding, which is limited by the token length of 512, it is more necessary to use a better method to make the model encode longer sequences.

% In this paper, we use text instead of id to represent items. With the encoding ability of the pretrained language model, we use the pretrained language model to encode sequences and items, and convert the recommendation task into dense vector retrieval task. At the same time, we refer to the practice of FID, and we realize long sequence modeling by separately encoding the long sequence, and then fusing it at the decode end.

\section{Related Work}
Sequential recommendation systems attempt to model user behaviors according to the user-item interaction history. Early work~\cite{He2016fusing,Rendle2010Factor} adopts Markov Chain assumption and Matrix Factorization Methods to recommend items. Existing sequential recommendation systems employ Convolutional Neural Networks (CNNs)~\cite{Jiaxi2018Personalized,Fajie2018asc,yan2019cosrec}, Recurrent Neural Networks (RNNs)~\cite{Hidasi2015session,liu2016context,donkers2017sequential}, Graph Neural Networks (GNNs)~\cite{Chang2021graph,zhang2022dynamic,cai2023lightgcl} or the self-attention architecture~\cite{Wang2018self,Sun2019Bert4rec,du2023frequency} to capture the dependencies among items in the user-item interaction sequences and predict the next item.

These recommendation systems represent items with only their ids and learn relevance among items using user-item interactions, even with Transformers. Bert4Rec~\cite{Sun2019Bert4rec} represents items with randomly initialized embeddings and pretrains self-attention heads~\cite{Vaswani2017attention} and item embeddings by mask language modeling~\cite{devlin2018bert}. P5~\cite{Geng0FGZ22} represents items using utterances of their ids. It converts multiple recommendation tasks as seq2seq tasks and continuously trains T5 model~\cite{raffel2020exploring} to generate the text representations of item ids that will be interacted with users in the next step. 

In addition to representing items with identifiers, using item attributes shows convincing recommendation results by modeling item dependency and user-item relevance through the side information~\cite{Liu2021nova,Xie2022DIF-SR,Zhou2020s3}. Existing work incorporates item attributes and focuses more on fully using additional information to identify the user intentions from the attributes of user-interacted items~\cite{ZhangZLSXWLZ19}. Moreover, some work~\cite{Liu2021nova,Xie2022DIF-SR} further builds sophisticated attention mechanisms to denoise item attributes and focuses more on fusing side information to enhance the representations of items. \citet{Zhou2020s3} also design an additional pretraining task to learn the embeddings for item attributes and identifiers, aiming to more effectively capture the correlation among item identifiers, attributes, and users. Nevertheless, these sequential recommendation systems often rely on randomly initialized embeddings to represent item attributes and identifiers, making it challenging to leverage the advantages of pretrained language models for enhancing sequential recommendation tasks.

Instead of using randomly initialized item embeddings, fully text-based item modeling recently emerges in recommendation systems~\cite{Ding2021zero,Hou2022uni,li2023text,yuan2023go}. It has proven that these randomly initialized item embeddings heavily depend on training with sufficient user-item interactions and usually face the cold start problem during representing long-tail items~\cite{lam2008addressing,schein2002methods,pan2019warm,zhang2021language}. \citet{li2023text} explore the potential of ID-free style recommendation modeling. They propose a full text matching based method to alleviate the cold start problem and benefit the cross-domain recommendations. They employ pretrained language models to provide the token embeddings of the text sequences of items for downstream recommendation models. \citet{yuan2023go} further discuss the advantages of id-based and modality-based recommendations, which represent items by randomly initialized embeddings and encoding multi-modal side information with pretrained language models, respectively. They confirm that id-based recommendation is usually effective when the user-item training signal is sufficient and modality-based recommendation can alleviate the cold start problem. 
However, \citet{yuan2023go} do not fully integrate both id-based and modality-based item representations, which could potentially contribute to further enhancing recommendation performance.%For modeling the text relevance between users and items, dense retriever~\cite{karpukhin2020dense,xiong2020dense,luan2020sparse,Yu2021FewShotCD} provides some successful attempts for text matching. Thus, we can encode user-item interactions and items using pretrained language models, map them in an embedding space for retrieval, and regard the user-item interaction sequences as queries to retrieve items. The encoders can be contrastively trained with in-batch negatives~\cite{karpukhin2020dense} and hard negatives~\cite{xiong2020approximate} to learn tailored representations of users and items.

% Modeling long-term user-item interaction can improve model performance~\cite{pi2019practice}. However, using side information of item to verbalize user-item interactions makes the text sequences long and challenges the long sequence processing ability of existing language models~\cite{devlin2018bert,Vaswani2017attention,raffel2020exploring}. Some work~\cite{qin2020user,cao2022sampling} tries to filter out unrelated items from the interaction history to avoid noise and reduce the user-item interaction length. Nevertheless, it is inevitable to lose some user preference information to model user intentions. Another way to model long text sequence is to make the attention sparse and reduce the self-attention computations~\cite{dai2019transformer,beltagy2020longformer,kitaev2020reformer,wang2020linformer}.
% \begin{figure}[t]
% \centering
% \includegraphics[width=\linewidth]{image/model.pdf}
% \caption{The Architecture of Dense Retrieval based Sequential Recommendation (SREDR).}
% \label{fig:model}
% \end{figure}
\begin{figure*}[ht]
\centering
\includegraphics[width=\linewidth]{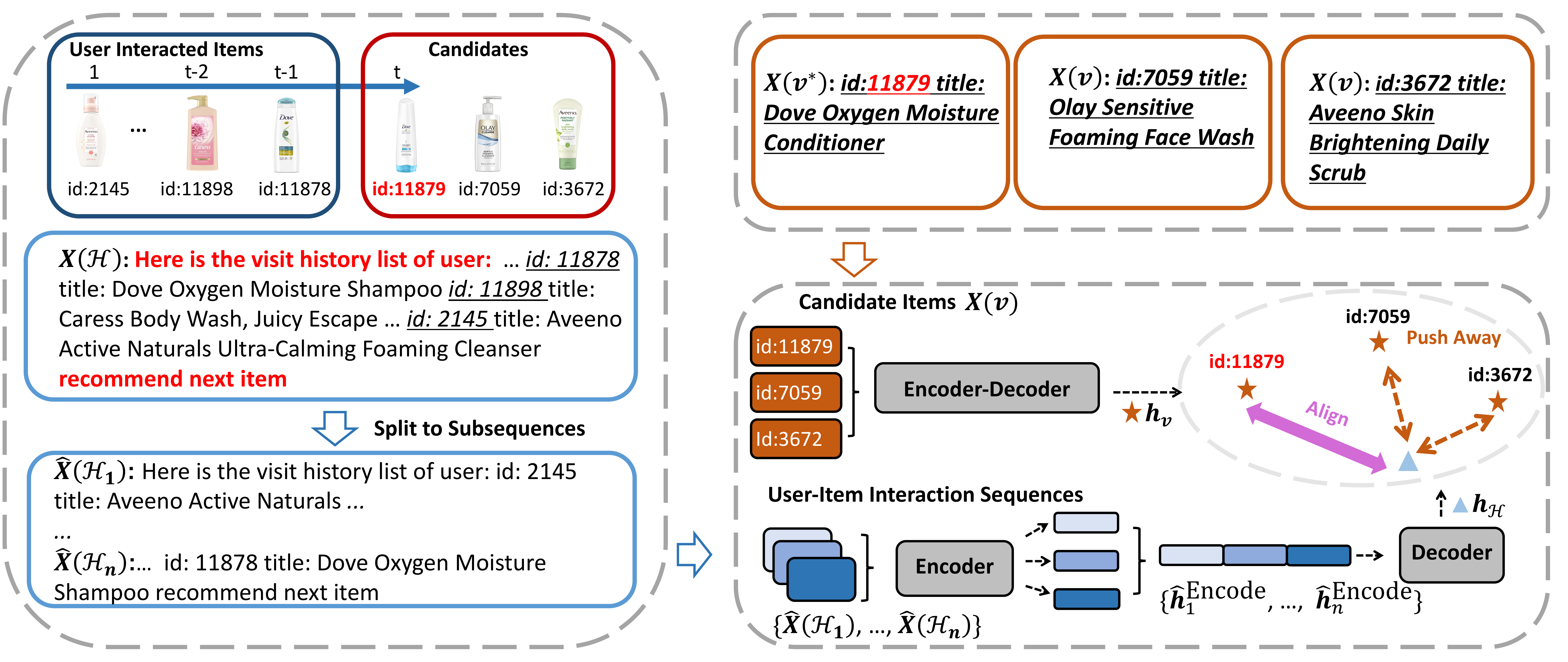}
\caption{The Framework of \textbf{T}ext m\textbf{A}tching based \textbf{S}equen\textbf{T}ial r\textbf{E}commendation (TASTE).}
\label{fig:model}
\end{figure*}

Modeling long-term user-item interaction can better characterize the user behaviors and improve the recommendation accuracy~\cite{pi2019practice}. However, using side information of items to verbalize user-item interactions makes the text sequences long and challenges existing language models~\cite{devlin2018bert,Vaswani2017attention,raffel2020exploring}. The work~\cite{qin2020user,cao2022sampling} tries to filter out unrelated items from the user-item interaction history to reduce the user-item interaction length. Nevertheless, it is inevitable to lose some user preference information during modeling user intentions. To model long text sequences, the work in other research directions shows some successful attempts. Some of them divide the long sequences into segments and separately encode them using pretrained language models and then fuse them~\cite{gong2020recurrent,li2020parade,fid}. The others make the attention sparse~\cite{child2019generating,beltagy2020longformer,kitaev2020reformer,zaheer2020big} or adopt low rank approximate~\cite{katharopoulos2020transformers,wang2020linformer,choromanski2020rethinking} to reduce the self-attention computations. Different from the work, TASTE separates the user-item interaction into different sessions and employs the fusion-in-decoder architecture~\cite{fid} to model the long text sequences.

% \citet{ivgi2023efficient} divides long documents into chunks and encodes them independently, and realizes the aggregation of chunks with the help of the Fusion in Decoder module~\cite{fid}.

\section{Methodology}
This section introduces our \textbf{T}ext m\textbf{A}tching based \textbf{S}equen\textbf{T}ial r\textbf{E}com-mendation (TASTE) model, which is illustrated in Figure~\ref{fig:model}. We first describe how to verbalize users and items and then model the sequential recommendation using text matching (Sec.~\ref{model:dr}). Finally, TASTE proposes an attention sparsity method to encode long text sequences of user-item interactions (Sec.~\ref{model:sparse}). 
% Different from previous work~\cite{Wang2018self,Sun2019Bert4rec}, our TASTE model targets generalizing items and user-item interaction history using ids and attributes of items and then encodes them in one embedding space for recommending items (Sec.~\ref{model:dr}). To model the long text sequence of user-item interactions, TASTE proposes an attention sparsity method to break the boundary of long text sequence modeling (Sec.~\ref{model:sparse}). 

\subsection{Text Matching based Sequential Recommendation Modeling}~\label{model:dr}
Given a user-item interaction history $\mathcal{H}=\{v_1,v_2,\dots,v_{t-1} \}$, the sequential recommendation task aims to recommend the item $v_t$ to satisfy the user's needs at $t$-th time. TASTE encodes $\mathcal{H}$ and item $v$ using pretrained language model T5~\cite{raffel2020exploring} and models the relevance between users and items by matching their text representations. 

\textbf{Text Representation.} For each item $v$, we can verbalize it using item ids and $k$ item attributes $\text{<Attr>}$ using the following template:
\begin{equation}\label{eq:item}
\small
X(v)= \text{id:} v(\text{id}) ... \text{<Attr>}_k: v(\text{<Attr>}_k),
\end{equation}
where $\text{<Attr>}_k$ is the name of attribute. $v(\text{id})$ and $v(\text{<Attr>}_i)$ are the text descriptions of item identifier and $i$-th attribute of item $v$. In this case, one item in Yelp can be verbalized: ``id: 5908 title: DoMazing address: 6659 S Las Vegas Blvd, Ste B-101 Las Vegas NV''. Here $v(\text{id})$ is a kind of id prompt to help language models to capture matching signals among users and items, which is beyond the descriptions of item attributes. 

Then we verbalize the user-item interaction history $\mathcal{H}$ with the template: ``Here is the visit history list of user: $X(\mathcal{H})$ recommend next item''. It demonstrates the definition of sequential recommendation and helps pretrained language models better characterize model behaviors~\cite{Geng0FGZ22,chung2022scaling}. $X(\mathcal{H})$ is the concatenation of verbalized items $\{v_1,v_2,\dots,v_{t-1} \}$ that are in $\mathcal{H}$:
\begin{equation}\label{eq:seqtext}
\small
X(\mathcal{H})=  X(v_{t-1}); ...; X(v_1),
\end{equation}
where $;$ denotes the concatenation operation. We reverse the item sequences for truncating more previously interacted items.

\textbf{Encoding.} We use T5~\cite{raffel2020exploring} to encode the user-item interaction history $\mathcal{H}$ and item $v$ as embeddings $h_\mathcal{H}$ and $h_v$, using the representation of the first input token from the T5 decoder~\cite{sentencet5}:
\begin{equation}\label{eq:encode}
\small
h_{\mathcal{H}} = \text{T5}(X(\mathcal{H})); h_{v} = \text{T5}(X(v)).
\end{equation}
The relevance between user-item interaction history $\mathcal{H}$ and item $v$ can be calculated using the ranking probability $P(v|\mathcal{H})$ between the encoded representations $h_\mathcal{H}$ and $h_v$: 
\begin{equation}
\small
P(v|\mathcal{H}) = \text{Softmax}_{v} (h_{\mathcal{H}} \cdot h_{v}),
\end{equation}
where $\cdot$ is the dot product operation. 

\textbf{Training.} We optimize the model parameters using the following loss function:
\begin{equation}\label{eq:cl}
\small
\mathcal{L} = \text{CrossEntropy} (P(v|\mathcal{H}), v^*),
\end{equation}
where $v^*$ denotes the ground truth item that is interacted by the user at the $t$-th time. We use in-batch negatives and randomly sampled negatives~\cite{rendle2012bpr,diaz2012real,he2017neural} to contrastively train models.

 %For each user, the items that interacted with are sorted in chronological order $S_u=\{v^u_1,v^u_2,\dots,v^u_t \}$ and defined as user history. Therefore, for the sequential recommendation task, the user's previous t-1 time history interaction is given, and the goal is to predict the items he interacted with at time t:$ P(v^u_t|S^{t-1}_u)$.

%For the fusion of side information, you only need to add the corresponding attributes after the tiitle of the item.

%For the processing of long text, we refer to the practice of FID. For the long text is divided into two cases, the first is to expand the user history on the basis of the original sequence, if the user interacts with 10 items and the original sequence can only store 6 items, then we will store the text of the remaining 4 items Also independently encode, and then calculate the attention for the spliced representation when decoding.

\subsection{Long Text Encoding for User-Item Interactions with Attention Sparsity}~\label{model:sparse}
In real-world scenarios, purchase or visit histories typically involve long-term interactions. The longer interaction usually contains more information to better model user behaviors and achieve more accurate recommendation results~\cite{pi2019practice}. Instead of only using item ids to model user behavior~\cite{Geng0FGZ22}, TASTE verbalizes the items in user-item interactions (Eq.~\ref{eq:seqtext}) and makes the text utterance $X(\mathcal{H})$ longer, which challenges the long text processing ability of pretrained language models due to the max length boundary~\cite{devlin2018bert,Vaswani2017attention,raffel2020exploring}.

As shown in the bottom part of Figure~\ref{fig:model}, our attention sparsity mechanism learns session-based user behaviors by employing a fusion-in-decoder architecture~\cite{fid}. We first split the text sequence $X(\mathcal{H})$ of user-item interaction history into $n$ sub-sequences $\hat{X}(\mathcal{H}) = \{\hat{X}(\mathcal{H})_1, ..., \hat{X}(\mathcal{H})_n\}$. $\hat{X}(\mathcal{H})_i$ contains $m$ tokens, which is regarded as a user-item interaction session. It reflects the user preferences of a specific period and can be characterized using a set of items that are interacted with users in the session~\cite{feng2019deep,wu2019session,li2017neural}. We use T5-Encoder to independently encode these subsequences:
\begin{equation}\label{eq:sparse}
\small
\hat{h}^\text{Encode}_i = \text{T5-Encoder} (\hat{X}(\mathcal{H})_i).
\end{equation}
Then we concatenate $\hat{h}^\text{Encode} = \{\hat{h}^\text{Encode}_1; \hat{h}^\text{Encode}_2; ...; \hat{h}^\text{Encode}_n\}$ as the hidden state of the user-item interaction history $\mathcal{H}$. It eliminates attention computations among different sessions within Transformer encoder modules, thereby reducing the encoding computational complexity from $O(n^2)$ to $O(n)$. This approach also facilitates the encoding of longer text sequences of user-item interactions.

Finally, we can get the representation of user-item interaction by feeding the sparsely encoded user-item interaction sequences to the decoder module of T5: 
\begin{equation}
\small
h_{\mathcal{H}} = \text{T5-Decoder} (\hat{h}^\text{Encode}, h^\text{Decode}_0),
\end{equation}
where $h^\text{Decoder}_0$ is the token embedding of the first input token of the decoder, which is the same as Eq.~\ref{eq:encode}. T5-Decoder uses the cross-attention mechanism to reweight different user-item interaction sessions and model the user behaviors by capturing text-matching signals from all tokens in the verbalized user-item interactions.

%\subsection{Text Combined with Item Identifier Modeling}~\label{model:itemid}
%Different from traditional modeling methods~\cite{Wang2018self}, TASTE uses the name and attributes of each item to verbalize it, thereby breaking the limitation of explicit ID modeling. However, it relies on the quality of the text which item names that are not clear or too similar may pose obstacles. To address this potential limitation, we propose TASTE+, where we extend TASTE by treating item ids as side information like attributes. Therefore the item can be represented by the following text sequence:
%\begin{equation}\label{eq:itemid}
%\small
%X(v)= \text{id}: v(\text{id}),\text{name}: v(\text{name}), ..., \text{<Attr>}_k: v(\text{<Attr>}_k),
%\end{equation}
%where add the id as part of the text on the basis of Eq.~\ref{eq:item}. For the user-item interaction history $\mathcal{H}$, we use the same representation as Eq.~\ref{eq:seqtext}.
%\input{table/datastatistic.tex}

% \begin{table}[t]
% \begin{center}
% \small
% \begin{tabular}{l | rr | r r r}
% \hline 
% \textbf{Dataset} &\textbf{\#User} & \textbf{\#Item} & \textbf{Train} & \textbf{Dev} & \textbf{Test}  \\ \hline
% Beauty & 22,363 & 12,101 & 131,413 & 22,363 & 22,363 \\
% Yelp & 30,499 & 20,068 & 225,685 & 30,499 & 30,499 \\
% Sports & 35,598 & 18,357 & 189,543 & 35,598 & 35,598 \\
% Toys & 19,412 & 11,924 & 109,361 & 19,412 & 19,412 \\
% \hline
% \end{tabular}
% \end{center}
% \caption{\label{tab:split}Data Statistics.}
% \end{table}

\begin{table}[t]
\begin{center}
\small
\caption{\label{tab:split}Statistics of Preprocessed Datasets.}
\resizebox{0.98\linewidth}{!}{
\begin{tabular}{l | r r r | r r r}
\hline 
\multirow{2}{*}{\textbf{Dataset}} &  \multicolumn{3}{c|}{\textbf{Data Information}} &  \multicolumn{3}{c}{\textbf{Split}}\\
&\textbf{\#Users} & \textbf{\#Items} & \textbf{\#Actions} & \textbf{Train} & \textbf{Dev} & \textbf{Test}  \\ \hline
Beauty & 22,363 & 12,101 & 198,502 & 131,413 & 22,363 & 22,363 \\
Yelp & 30,499 & 20,068 & 317,182 & 225,685 & 30,499 & 30,499 \\
Sports & 35,598 & 18,357 & 296,337 & 189,543 & 35,598 & 35,598 \\
Toys & 19,412 & 11,924 & 167,597 & 109,361 & 19,412 & 19,412 \\
\hline
\end{tabular}}
\end{center}
\end{table}
\section{Experimental Methodology}\label{sec:exp}
This section describes datasets, evaluation metrics, baselines, and implementation details in our experiments.

\textbf{Dataset.}
Four sequential recommendation datasets are used in our experiments, including Yelp\footnote{\url{https://www.yelp.com/dataset}}, Amazon Beauty, Sports and Toys~\cite{mcauley2015image}, which aim to recommend locations and Amazon products\footnote{\url{http://jmcauley.ucsd.edu/data/amazon/}} to satisfy users.  All data statistics are shown in Table~\ref{tab:split}.

We use Recbole~\cite{zhao2021recbole} to process all datasets and keep all experiment settings the same as previous work~\cite{Xie2022DIF-SR,Zhou2020s3}, which allows us to directly compare TASTE with baseline models~\cite{Xie2022DIF-SR}. For all datasets, we filter out the items and users that have less than five user-item interaction times in the whole datasets and treat all user-item interactions as implicit feedback~\cite{Xie2022DIF-SR,chen2018sequential,Sun2019Bert4rec,Zhou2020s3}. After data processing, each example can be a user-item interaction sequence $\mathcal{H}=\{v_1,v_2,\dots,v_{T} \}$. Then we use the leave-one-out evaluation strategy~\cite{Xie2022DIF-SR,chen2018sequential,Sun2019Bert4rec,Zhou2020s3}, and separate the processed datasets into training, development, and testing sets. We construct the testing and development sets by using $v_{1,...,T-1}$ to predict $v_T$ and using $v_{1,...,T-2}$ to predict $v_{T-1}$, respectively.
For the training set, we follow~\citet{zhao2021recbole} and ~\citet{Xie2022DIF-SR} to use interaction history $v_{1,...,i-1}$ to predict $v_i$, where $ 1<i<T-1$.
%We follow previous work~\cite{Xie2022DIF-SR} to process datasets. For the user-item interaction sequence $\mathcal{H}=\{v_1,v_2,\dots,v_{T} \}$, we construct the testing and development sets by using $v_{1,...,T-1}$ to predict $v_T$ and using $v_{1,...,T-2}$ to predict $v_{T-1}$, respectively.
%For the training set, we follow~\citet{zhao2021recbole} to use interaction history $v_{1,...,i-1}$ to predict $v_i$ where $ 1<i<T-1$.  We showcase more experimental details of dataset processing and data statistics in the Appendix~\ref{app:datapro}.

\textbf{Evaluation Metrics.}
We utilize the same evaluation metrics as DIF-SR~\cite{Xie2022DIF-SR} and use Recall@10/20 and NDCG@10/20 to evaluate the recommendation performance of different models. Statistic significances are tested by permutation test with P$<0.05$.

During evaluating models, we follow DIF-SR~\cite{Xie2022DIF-SR} and employ a full ranking testing scenario~\cite{dallmann2021case,krichene2022sampled}. Some work evaluates model performance on a small item subset by randomly sampling or sampling items according to item popularity, making the evaluation results inconsistent of the same model. Instead of reranking items in the sampled subset, some work~\cite{krichene2022sampled,dallmann2021case} builds a more realistic recommendation evaluation setting by ranking all items and choosing the top-ranked items as recommendation results.

\textbf{Baselines.}
Following our main baseline model~\cite{Xie2022DIF-SR}, we compare TASTE with several widely used sequential recommendation models.
GRU4Rec~\cite{Hidasi2015session} uses RNN to model user-item interaction sequences for recommendation. SASRec~\cite{Wang2018self} and Bert4Rec~\cite{Sun2019Bert4rec} employ the self-attention mechanism to capture user preferences from user-item interaction sequences. To better capture relevance among attributes, items, and users, S$^3$Rec~\cite{Zhou2020s3} comes up with four self-supervised methods to pretrain self-attention modules. ICAI-SR~\cite{yuan2021icai} is compared, which proposes a heterogeneous graph to represent the relations between items and attributes to model item relevance. Besides, NOVA~\cite{Liu2021nova} is also compared, which builds attention modules to incorporate item attributes as the side information. DIF-SR~\cite{Xie2022DIF-SR} is the previous state-of-the-art, which builds a non-invasive attention mechanism to fuse information from item attributes during modeling user behaviors.
%ICAI-SR~\cite{yuan2021icai} is compared, which proposes a heterogeneous graph to represent the relations between items and attributes to model item relevance. Besides, NOVA~\cite{Liu2021nova} and DIF-SR~\cite{Xie2022DIF-SR} are also compared, which build non-invasive attention mechanisms to fuse attributes and model side information for recommendation. 

Besides, we implement a T5-ID model, which encodes user-item interaction sequences to generate the identifier of the next item~\cite{Geng0FGZ22}. We follow the previous dense retrieval model, DPR~\cite{karpukhin2020dense}, to implement T5-DPR, which regards the text sequences of user-item interaction history and items as queries and documents, encodes them with T5 and is trained with in-batch sampled negatives.

\textbf{Implementation Details.}
Different from previous work~\cite{Wang2018self,Sun2019Bert4rec,Xie2022DIF-SR}, we represent items using full-text sequences (Eq.~\ref{eq:item}) instead of randomly initialized item embeddings. In our experiments, we use names and addresses to represent locations in the Yelp dataset. And we only use product names to represent shopping products in Amazon Product datasets, because the product name usually contains the item attributes, such as ``\textit{WAWO 15 Color Professional Makeup Eyeshadow Camouflage Facial Concealer Neutral Palette}'' contains both category and brand.

Our models are implemented with OpenMatch~\cite{liu2021openmatch,yu2023openmatch}. In our experiments, we truncate the text representations of items by 32 tokens and set the max length of the text representation of user-item interactions to 512. The user-item interaction sequence is split into two subsequences to make the attention sparse in T5-encoder (Eq.~\ref{eq:sparse}). TASTE is initialized with the T5-base checkpoint from huggingface transformers~\cite{wolf2019huggingface}. During training, we use Adam optimizer and set learning rate=1e-4, warm up proportion=0.1, and batch size=8. Besides, we use in-batch negatives and randomly sampled negatives to optimize TASTE. We randomly sample 9 negative items for each training example and in-batch train TASTE model.

\begin{table*}
\small
\centering
\caption{\label{tab:overall}Overall Performance. We keep the same experimental settings and report the scores of baselines from previous work~\cite{Xie2022DIF-SR}. Underlined scores are the highest results of baselines. ${\dagger}$, ${\mathsection}$, ${\ddagger}$ indicate statistically significant improvements over $\text{DIF-SR}^{\dagger}$, $\text{T5-DPR}^{\mathsection}$ and $\text{T5-ID}^{\ddagger}$, respectively. We also show relative improvements over DIF-SR.}
\resizebox{\textwidth}{!}{
\begin{tabular}{l|c| c   c  c  c c c   c  | c c |    c r    }
\hline 
\textbf{Dataset} & \textbf{Metrics} & \textbf{GRU4Rec}  &  \textbf{Bert4Rec}  &  \textbf{SASRec}  & \textbf{S$^3$Rec} &  \textbf{NOVA}  & \textbf{ICAI-SR}  & \textbf{DIF-SR} & \textbf{T5-DPR} & \textbf{T5-ID} &  \multicolumn{2}{c}{\textbf{TASTE}}\\ 
\hline
\multirow{4}{*}{Beauty} & Recall@10 & 0.0530 & 0.0529 &  0.0828  &0.0868 &  0.0887  &0.0879 & \uline{0.0908} &  0.0716 & 0.0785$\text{}^{\mathsection}$ & \textbf{0.1030}$\text{}^{\dagger \ddagger \mathsection}$ & 13.44\%  \\
& Recall@20  & 0.0839  & 0.0815  & 0.1197 & 0.1236 & 0.1237 & 0.1231 & \uline{0.1284} &  0.1082 & 0.1138$\text{}^{\mathsection}$  & \textbf{0.1550}$\text{}^{\dagger \ddagger \mathsection}$ & 20.72\%  \\
& NDCG@10  & 0.0266 & 0.0237 & 0.0371 & 0.0439 & 0.0439 & 0.0439 & \uline{0.0446} &  0.0345 & 0.0440$\text{}^{\mathsection}$ &  \textbf{0.0517}$\text{}^{\dagger \ddagger \mathsection}$ & 15.92\%  \\
& NDCG@20  & 0.0344  & 0.0309  & 0.0464 & 0.0531 & 0.0527 & 0.0528 & \uline{0.0541} &  0.0437 & 0.0529$\text{}^{\mathsection}$ & \textbf{0.0649}$\text{}^{\dagger \ddagger \mathsection}$ & 19.96\%  \\
\hline
\multirow{4}{*}{Sports} & Recall@10  & 0.0312 &0.0295 &0.0526 & 0.0517 &0.0534 & 0.0527 & \uline{0.0556} &  0.0329 & 0.0464$\text{}^{\mathsection}$ & \textbf{0.0633}$\text{}^{\dagger \ddagger \mathsection}$ & 13.85\% \\
& Recall@20 & 0.0482 &0.0465 &0.0773 &0.0758 &0.0759 &0.0762 & \uline{0.0800} &  0.0554 & 0.0689$\text{}^{\mathsection}$ & \textbf{0.0964}$\text{}^{\dagger \ddagger \mathsection}$ & 20.50\% \\
& NDCG@10  & 0.0157 & 0.0130 &0.0233 &0.0249 &0.0250 &0.0243 & \uline{0.0264} &  0.0159 & 0.0252$\text{}^{\mathsection}$ & \textbf{0.0338}$\text{}^{\dagger \ddagger \mathsection}$ & 28.03\%  \\
& NDCG@20 & 0.0200 & 0.0173 &0.0295 & 0.0310 &0.0307 & 0.0302 & \uline{0.0325} &  0.0216 & 0.0308$\text{}^{\mathsection}$  & \textbf{0.0421}$\text{}^{\dagger \ddagger \mathsection}$ & 29.54\%  \\
\hline
\multirow{4}{*}{Toys} & Recall@10  & 0.0370  & 0.0533 &  0.0831  &0.0967 &  0.0978  &0.0972 & \uline{0.1013} &  0.0805$\text{}^{\ddagger}$ & 0.0754 & \textbf{0.1232}$\text{}^{\dagger \ddagger \mathsection}$ & 21.62\% \\
& Recall@20  & 0.0588  &  0.0787  & 0.1168 & 0.1349 & 0.1322 & 0.1303 & \uline{0.1382} &  0.1243$\text{}^{\ddagger}$ & 0.1065 & \textbf{0.1789}$\text{}^{\dagger \ddagger \mathsection}$ & 29.45\% \\
& NDCG@10 & 0.0184 &  0.0234 & 0.0375 & 0.0475 & 0.0480 & 0.0478  & \uline{0.0504} &  0.0375 & 0.0421$\text{}^{\mathsection}$ & \textbf{0.0640}$\text{}^{\dagger \ddagger \mathsection}$ & 26.98\% \\
& NDCG@20 & 0.0239  & 0.0297  & 0.0460 & 0.0571 & 0.0567 & 0.0561 & \uline{0.0597}&  0.0485 & 0.0499 & \textbf{0.0780}$\text{}^{\dagger \ddagger \mathsection}$ & 30.65\% \\
\hline
\multirow{4}{*}{Yelp} & Recall@10  & 0.0361 & 0.0524 &  0.0650  &0.0589 &  0.0681  &0.0663  & \uline{0.0698} &  0.0460 & 0.0450 & \textbf{0.0738}$\text{}^{\dagger \ddagger \mathsection}$ & 5.73\% \\
& Recall@20 & 0.0592  & 0.0756  & 0.0928 & 0.0902 & 0.0964 & 0.0940 & \uline{0.1003} &  0.0740 & 0.0745 & \textbf{0.1156}$\text{}^{\dagger \ddagger \mathsection}$ & 15.25\% \\
& NDCG@10  & 0.0184 & 0.0327 & 0.0401 & 0.0338 & 0.0412 & 0.0400 & \uline{\textbf{0.0419}} &  0.0250$\text{}^{\ddagger}$ & 0.0235 & 0.0397$\text{}^{\ddagger \mathsection}$ & -5.25\%  \\
& NDCG@20 & 0.0243  & 0.0385  & 0.0471 & 0.0416 & 0.0483 & 0.0470 & \uline{0.0496} &  0.0320 & 0.0309 & \textbf{0.0502}$\text{}^{\ddagger \mathsection}$ & 1.21\% \\
 \hline

\end{tabular}
}

\end{table*}

\section{Evaluation Result}
In this section, we first evaluate the recommendation performance of TASTE and conduct ablation studies. Then we study the effectiveness of different item verbalization methods, the ability of TASTE in reducing the popularity bias, the advantages of TASTE in representing long-tail items, and modeling user behaviors using longer user-item interactions. Finally, we present several case studies.

\subsection{Overall Performance}
The recommendation performance of TASTE is shown in Table~\ref{tab:overall}. Overall, TASTE significantly outperforms baseline models on all datasets by achieving 18\% improvements. %Such improvements illustrate the effectiveness of TASTE in recommending the next items for users.

Compared with item id embedding based recommendation models, \textit{e.g.} Bert4Rec, TASTE almost doubles its recommendation performance, thriving on modeling relevance between users and items through text-matching signals. TASTE also outperforms the previous state-of-the-art recommendation model, DIF-SR, which leverages item attributes as side information to help better learn user behaviors from user-item interactions beyond item identifiers. It represents item attributes as embeddings and focuses more on decoupling and fusing the side information in item representations. Instead of designing sophisticated architectures for fusing side information, TASTE employs a general template to verbalize items and users, leveraging pretrained attention heads of T5 to match the textual representations of users and items. It demonstrates the direct advantages of utilizing pretrained language models for recommendation systems. %On the other hand, TASTE also achieves consistent improvements over T5 and T5-DPR~\cite{karpukhin2020dense}, showing the advances of incorporating both id and text. Then we conduct the following experiments to explore the effectiveness of TASTE in generalizing items and modeling user behaviors.

% setting:inbatch:batch size=16 seq maxlen=256,item maxlen=32. popular:All items that users have interacted with in the statistical data set, sorted from high to low. Keep top500, and select 100 from top500 as negative samples for each sample.
\begin{table*}
\centering
\small
\caption{Ablation Study. We show the effectiveness of different negative sampling strategies during training TASTE, including Popular Negs, Hard Negs, and Random Negs. The popular negatives are selected from the top-500 items that are more frequently interacted with users. The Longer History uses our attention sparsity strategy for modeling longer user-item interactions.\label{tab:ablation}}
\begin{tabular}{l|c| c|  c  c   c  c  |  c  }
\hline 
\multirow{2}{*}{\textbf{Dataset}} & \multirow{2}{*}{\textbf{Metrics}} &  \textbf{TASTE w/o Prompt} &  \multicolumn{4}{c|}{\textbf{TASTE}} & \textbf{TASTE (Rand Negs)} \\
& &\textbf{(Inbatch)} & \textbf{Inbatch}  & \textbf{Popular Negs}  & \textbf{ANCE} & \textbf{Random Negs} & \textbf{w/ Longer History} \\ 
\hline
\multirow{4}{*}{Yelp} & Recall@10 & 0.0441 &  0.0460  &0.0351 & 0.0324 & 0.0726& \textbf{0.0733}  \\
& Recall@20  & 0.0714  & 0.0740 & 0.0550 & 0.0519 & 0.1131& \textbf{0.1150}  \\
& NDCG@10  & 0.0235 & 0.0250 & 0.0186& 0.0173 & 0.0388& \textbf{0.0393}  \\
& NDCG@20  &0.0304  & 0.0320 & 0.0236 & 0.0222 & 0.0489& \textbf{0.0498}  \\
\hline
\multirow{4}{*}{Sports} & Recall@10   &0.0327 &0.0329 & 0.0284 & 0.0354 & 0.0510& \textbf{0.0545} \\
& Recall@20 &0.0550 &0.0554 &0.0476 & 0.0550 & 0.0812& \textbf{0.0851} \\
& NDCG@10  &0.0155 &0.0159 &0.0135 & 0.0182 & 0.0249& 
 \textbf{0.0270}  \\
& NDCG@20 &0.0211 &0.0216 & 0.0184 & 0.0232 & 0.0325& \textbf{0.0346}  \\
\hline
\multirow{4}{*}{Beauty} & Recall@10  & 0.0688 & 0.0716 &  0.0601  &0.0750 & 0.0921 & \textbf{0.0935} \\
& Recall@20  & 0.1085  & 0.1082 & 0.0950 & 0.1147 & 0.1401& \textbf{0.1441} \\
& NDCG@10 & 0.0335 & 0.0345 & 0.0294  & 0.0368 & 0.0444& \textbf{0.0445} \\
& NDCG@20 &0.0435  & 0.0437 & 0.0382 & 0.0468 & 0.0565& \textbf{0.0573} \\
\hline
\multirow{4}{*}{Toys} & Recall@10  & 0.0777 &  0.0805  &0.0678  & 0.0878 & 0.1032& \textbf{0.1056} \\
& Recall@20 & 0.1222  & 0.1243 & 0.1065 & 0.1297 & 0.1577& \textbf{0.1594} \\
& NDCG@10  & 0.0369 & 0.0375 & 0.0326 & 0.0426 & 0.0488& \textbf{0.0508}  \\
& NDCG@20 &0.0480  & 0.0485 & 0.0423 & 0.0532 & 0.0625& \textbf{0.0643} \\
 \hline

\end{tabular}

\end{table*}

\begin{table}[t]
\centering
\small

\caption{\label{tab:sparse}Performance of the Attention Sparsity Module of TASTE. We evaluate TASTE on Yelp by splitting user-item interactions into different sequences.}
\begin{tabular}{l|l|ccc}
\hline
 \textbf{Model} & \textbf{Seq Length} & \textbf{Recall@20} &  \textbf{NDCG@20} & \textbf{Memory}\\ \hline
\multirow{6}{*}{TASTE}&256 & 0.1076  & 0.0468 & 9.52GB   \\
&128$\times$2 seqs & 0.1056 & 0.0453 & 9.18GB   \\
&64$\times$4 seqs  & \textbf{0.1097} & \textbf{0.0472} & \textbf{9.05GB}   \\
\cline{2-5}
&512 & 0.1142  & 0.0499 & 13.09GB   \\
&256$\times$2 seqs & \textbf{0.1156}  & \textbf{0.0502} & 11.84GB   \\
&128$\times$4 seqs & 0.1090  & 0.0470 & \textbf{11.16GB}   \\ \hline

% 64$\times$4 seq  & 0.1052 & 0.0450 & 9.49GB   \\
% 128$\times$2 seq & 0.1132 & 0.0488 & 9.57GB   \\
% 256 & 0.1131  & 0.0489 & 9.90GB   \\
% \hline
% 128$\times$4 seq & 0.1145  & 0.0495 & 12.28GB   \\
% 256$\times$2 seq & 0.1150  & 0.0498 & 12.95GB   \\ 
% 512 & 0.1161  & 0.0503 & 14.29GB   \\ \hline
\end{tabular}
\end{table}

\subsection{Ablation Study}\label{sec:ablation}
In this experiment, we further explore the effectiveness of prompt modeling, different negative sampling strategies, and our long user-item interaction modeling method.

We first conduct several experiments by in-batch training TASTE with 9 additional negatives that are from popular sampling, random sampling, and hard negative sampling. The popular negatives are sampled according to the item appearance frequencies in all user-item interaction sequences. The items that are higher-frequently interacted with users means more ``popular'' items. During sampling popular negatives, we follow~\citet{Sun2019Bert4rec} to build the popular item set, which consists of 500 items that are more frequently interacted with users. The hard negatives are sampled from TASTE (Inbatch), which are more informative to avoid vanishing gradients~\cite{xiong2020approximate}. For each user, all user-interacted items are filtered out from the negative item set.
As shown in Table~\ref{tab:ablation}, using additional negatives that are sampled from popular items slightly decreases the recommendation performance of TASTE (Inbatch). The reasons may lie in that high-frequently interacted items indicate the general interests of users and can be easily interacted with different users. TASTE (ANCE) outperforms TASTE (Inbatch) while showing less effectiveness than TASTE (Rand Negs). This phenomenon illustrates that 
these hard negatives are also high-potentially interacted with users and are usually not the real negatives. Besides, benefiting from our attention sparsity method, TASTE achieves further improvements, demonstrating its effectiveness in characterizing user behaviors by modeling longer-term user-item interactions.
% Besides, with the help of our attention sparsity method, TASTE is further improved, demonstrating its effectiveness in learning more accurate representations of user behaviors by modeling longer-term user-item interactions.

Finally, we show the effectiveness of our attention sparsity method in Table~\ref{tab:sparse}. In our experiments, we maintain maximum text sequence lengths of 256 and 512 for user-item interaction history, splitting them into 4 or 2 subsequences. Then we evaluate the memory usage and recommendation performance of TASTE. Overall, by adding more subsequences or extending their lengths to model additional user-interacted items, TASTE achieves more accurate recommendation results. It demonstrates that modeling longer user-item interaction sequences can help to better characterize user behaviors~\cite{pi2019practice}. When the sequences are separated into different numbers of subsequences, TASTE reduces the GPU memory usage and achieves even slightly better recommendation performance. Our attention sparsity method has the ability to reduce the self-attention computations and potentially break the boundary of existing pretrained language models to model long user-item interactions, which is important in more realistic scenarios that have sufficient product purchase history and restaurant visiting history.

% setting: The first three sets of experiments and the final model keep the same setting. 
% The experiment of adding all auxiliary information, the maximum length of the sequence is extended to 512

\begin{table}[t]
\centering
\small
\caption{\label{tab:side}Effectiveness of Attribute Information. We conduct several experiments on Yelp by filling in the templates using names, addresses, and categories to represent items.}
\begin{tabular}{l| c | c}
\hline \textbf{Attribute Info} & \textbf{Recall@20} &  \textbf{NDCG@20}\\ \hline
T5-DPR w/ Item ID  & 0.0687 & 0.0313   \\\hline
+ Name & 0.0915 & 0.0415 \\
+ Name \& Category & 0.0909 & 0.0410\\
+ Name \& Address & 0.1057 & 0.0451\\
+ Name \& Address \& Category  & \textbf{0.1107} & \textbf{0.0475}\\
\hline
\end{tabular}
\end{table}

% \begin{table}[t]
% \centering
% \small
% \caption{\label{tab:side}Effectiveness of Attribute Information. We conduct several experiments on Yelp by filling in the templates using names, categories, and addresses to represent items.}
% \resizebox{0.48\textwidth}{!}{
% \begin{tabular}{l| c | c | c | c}
% \hline \textbf{Attribute Info} & \textbf{Recall@10} & \textbf{Recall@20} &  \textbf{NDCG@10} &  \textbf{NDCG@20}\\ \hline
% % T5-DPR w/ Item ID  & 0.0687 & 0.0313   \\\hline
% % + Name & 0.0915 & 0.0415 \\
% % + Name \& Category & 0.0923 & 0.0417\\
% % + Name \& Category \& Address & \textbf{0.1116} & \textbf{0.0480}\\
% T5-DPR w/ Item ID & 0.0592 & 0.0860 & 0.0334 & 0.0401   \\\hline
% % + Item ID & 0.0606 & 0.0915 & 0.0337 & 0.0415 \\
% + Name & 0.0578 & 0.0895 & 0.0320 & 0.0400 \\
% + Name \& Category & 0.0578 & 0.0895 & 0.0320 & 0.0400 \\
% + Name \& Address & \textbf{0.0726} & \textbf{0.1131} & \textbf{0.0388} & \textbf{0.0489}   \\ 
% + Name \& Category \& Address & \textbf{0.0726} & \textbf{0.1131} & \textbf{0.0388} & \textbf{0.0489}   \\ 
% %w/ Category & 0.0895  & 0.0400   \\
% %w/ Address & 0.1131 & 0.0489   \\
% %w/ Address \& Category & 0.1148 & 0.0496    \\ 
% %w/ Address \& Category \&ID & 0.1107  & 0.0475 \\
% \hline
% \end{tabular}}
% \end{table}

\subsection{Effectiveness of Item Verbalization Methods}
In this experiment, we show the recommendation effectiveness of TASTE by using different item verbalization methods. We first study the effectiveness of item attributes and then model user/item ids using different methods. Finally, the recommendation behaviors of different item modeling methods are further explored.

% \textbf{Side Information Modeling.} As shown in Table~\ref{tab:side}, we first show the effectiveness of item attributes in representing items by adding names, categories, and addresses of locations in the Yelp dataset step by step. We do not conduct experiments on Amazon Products, because the names of these products usually contain the information of other attributes. Our experiments showcase that all attributes improve model performance, especially the address of the location. They potentially indicate the staying areas and preferences of users and provide more text-matching signals to model the relevance between users and items. TASTE can be easily improved by filling these attributes in the prompt template, demonstrating its expandability in modeling side information.
\textbf{Side Information Modeling.} In Table~\ref{tab:side}, we illustrate the effectiveness of different item attributes in representing items in the Yelp dataset.
% In Table~\ref{tab:side}, we first illustrate the effectiveness of item attributes in representing items by using names, and then we add categories and addresses of locations in the Yelp dataset, respectively. 
We do not conduct experiments on Amazon Products, because the names of these products usually contain the information of other attributes. 
Our experimental results show distinct effects on the two item attributes. 
% Adding location information significantly enhances model performance, whereas adding category information results in a performance decline. 
The reason might be that location information potentially indicates user staying areas, offering more text-matching signals for user-item relevance modeling. However, the category information consists of discrete words like "Greek," "Mediterranean," and "Restaurants," rather than sentences that the model finds easier to understand. Additionally, some of these words are partially included within the name attributes. Therefore, modeling category information has an adverse effect.
% However, learning to match category information limits the diversity of user needs for restaurants. 
TASTE can be easily improved by filling attributes in the prompt template, demonstrating its expandability in modeling side information.

\begin{table}[t]
\begin{center}
\small
\caption{\label{tab:idmodeling} Recommendation Performance of Different Identifier Modeling Strategies. The evaluation results of two identifier modeling methods, Embed and Prompt, are shown. Embed randomly initializes the embeddings of item/user, while Prompt verbalizes the identifiers in the text space. TASTE only models the item identifiers using the Prompt method.}
\resizebox{0.48\textwidth}{!}{
\begin{tabular}{l|c|c|   c  c |  c   c }
\hline 
\multirow{2}{*}{\textbf{Dataset}} & \multirow{2}{*}{\textbf{Metrics}} &  \textbf{TASTE} & \multicolumn{2}{c|}{\textbf{w/ User ID}} &  \multicolumn{2}{c}{\textbf{w/ Item ID}}\\\cline{4-7}
& &  \textbf{w/o ID} &\textbf{Embed}  &  \textbf{Prompt}  & \textbf{Embed}  &  \textbf{Prompt}\\ 
\hline
\multirow{4}{*}{Beauty} & Recall@10 & 0.0935 & 0.0600 &  0.0910  &0.0743 & \textbf{0.1030}  \\
& Recall@20 & 0.1441  & 0.0982  & 0.1399 & 0.1138 & \textbf{0.1550}  \\
& NDCG@10 & 0.0445  & 0.0289 & 0.0437 & 0.0374 & \textbf{0.0517}  \\
& NDCG@20 & 0.0573  &0.0385  & 0.0560 & 0.0474 & \textbf{0.0649}  \\
\hline
\multirow{4}{*}{Sports} & Recall@10 & 0.0545  &0.0358 &0.0522 & 0.0427 & \textbf{0.0633} \\
& Recall@20 & 0.0851 &0.0581 &0.0817 &0.0661 & \textbf{0.0964} \\
& NDCG@10 & 0.0270  &0.0184 &0.0258 &0.0218 & \textbf{0.0338}  \\
& NDCG@20 & 0.0346 &0.0240 &0.0332 & 0.0277 & \textbf{0.0421}  \\
\hline
\multirow{4}{*}{Toys} & Recall@10 & 0.1056  & 0.0657 &  0.1021  &0.0750 & \textbf{0.1232} \\
& Recall@20 & 0.1594  & 0.1068  & 0.1552 & 0.1169 & \textbf{0.1789} \\
& NDCG@10 & 0.0508 & 0.0318 & 0.0486 & 0.0382  & \textbf{0.0640} \\
& NDCG@20 & 0.0643 &0.0422  & 0.0620 & 0.0488 & \textbf{0.0780} \\
\hline
\multirow{4}{*}{Yelp} & Recall@10 & 0.0733  & 0.0503 &  0.0731  &0.0492  & \textbf{0.0738} \\
& Recall@20 & 0.1150 & 0.0842  & 0.1150 & 0.0817  & \textbf{0.1156} \\
& NDCG@10 & 0.0393  & 0.0265 & \textbf{0.0401} & 0.0241 & 0.0397  \\
& NDCG@20 & 0.0498 &0.0350  & \textbf{0.0506} & 0.0322 & 0.0502 \\
 \hline
\end{tabular}}
\end{center}

\end{table}

\textbf{User/Item Identifier Modeling.} Then we further explore the recommendation effectiveness of TASTE using different user/item id modeling methods. The user/item ids are crucial to modeling the relevance and dependency among users and items~\cite{huang2013learning,yi2019sampling}. Here we explore two methods to model user/item ids to boost TASTE, including \texttt{Embed} and \texttt{Prompt}.
We follow~\citet{yuan2023go} to implement the \texttt{Embed} method. We first randomly initialize the id embeddings of users/items, encode the text utterances of item attributes and sum the id embedding and attribute embedding as the representations of users/items. Different from the \texttt{Embed} method, \texttt{Prompt} follows previous work~\cite{Geng0FGZ22}, verbalizes the item ids using texts (Eq.~\ref{eq:item}), and modifies the template by adding user ids, such as ``Here is the visit history list of user\_{}1401:''. The \texttt{Prompt} method regards user/item ids as a part of the templates in prompt learning.

The experimental results are shown in Table~\ref{tab:idmodeling}. The recommendation performance of TASTE is indeed increased with the help of item ids. It shows that the item ids probably provide additional text-matching signals to model dependency and relevance among items and users, making TASTE return more appropriate items as the recommendation results. On the other hand, user ids play different roles in product recommendation (Beauty, Sports, and Toys) and restaurant recommendation (Yelp). The main reason may lie in that shopping behaviors are less personalized and can be usually described by the shopping history. While modeling restaurant visiting behaviors needs more to memorize some characteristics of users, such as the taste, preferred cuisines, and active area of users.
Besides, the prompt-based modeling method (\texttt{Prompt}) outperforms the embedding-based method (\texttt{Embed}), which illustrates that pretrained language models have the ability to understand the user/item identifiers and establish relevance between users and items via identifiers. It further supports the motivation of TASTE, which fully uses the learned knowledge from pretrained language models to build sequential recommendation systems.

\begin{table}[t]
\begin{center}
\small
\caption{\label{tab:itemgen} Recommendation Behaviors with Different Item Verbalization Methods. Top-5 recommended items are used for evaluations. Bleu ($\uparrow$) and Dist ($\uparrow$) are used for evaluating the relevance and diversity of the text representations of items in recommendation results, respectively. Popular calculates the ratios of items, which are top-500 items that are more frequently interacted with users.}
% \resizebox{0.49\textwidth}{!}{
\begin{tabular}{l|l|   c  c  c }
\hline 
\textbf{Dataset} & \textbf{Metrics} &  \textbf{T5-ID}  &  \textbf{TASTE w/o ID}  & \textbf{TASTE}\\ 
\hline
\multirow{5}{*}{Beauty} & Dist-1 & \textbf{0.1456} &  0.1385  & 0.1398   \\
& Dist-2  & \textbf{0.5003}  & 0.4452 & 0.4641   \\
& Bleu-4  & 0.0267 & 0.0321 & \textbf{0.0322}   \\
& Recall-5 &0.0509 & 0.0540 & \textbf{0.0639} \\
& Popular & \textbf{52\%} & 29\% & 31\% \\
\hline
\multirow{4}{*}{Sports} & Dist-1 & \textbf{0.1740} &  0.1571  &0.1623   \\
& Dist-2  & \textbf{0.5668}  & 0.4917 & 0.5193   \\
& Bleu-4  & 0.0130 & 0.0153 & \textbf{0.0163}   \\
& Recall-5 & 0.0304 & 0.0326 & \textbf{0.0403} \\
& Popular & \textbf{62\%} & 36\% & 33\% \\
\hline
\multirow{4}{*}{Toys} & Dist-1 & \textbf{0.1728} &  0.1564  &0.1601   \\
& Dist-2  & \textbf{0.5268}  & 0.4493 & 0.4697   \\
& Bleu-4  & 0.0467 & 0.0566 & \textbf{0.0588}   \\
& Recall-5 & 0.0507 & 0.0636 & \textbf{0.0793} \\
& Popular & \textbf{38\%} & 29\% & 29\% \\
\hline
\multirow{4}{*}{Yelp} & Dist-1 & \textbf{0.1402} &  0.1363  & 0.1351   \\
& Dist-2  & \textbf{0.4367}  & 0.4184 & 0.4177   \\
& Bleu-4  & 0.0479 & \textbf{0.0542} & 0.0540   \\
& Recall-5 & 0.0276 & 0.0465 & \textbf{0.0469} \\
& Popular & \textbf{46\%} & 30\% & 30\% \\
 \hline
\end{tabular}
% }
\end{center}

\end{table}

\textbf{Evaluation on Recommendation Behaviors.} Finally, we explore the recommendation behaviors of TASTE using different item modeling methods. As shown in Table~\ref{tab:itemgen}, three models, T5-ID, TASTE w/o ID, and TASTE, are compared. T5-ID randomly initializes item embeddings and directly predicts the item ids. TASTE w/o ID and TASTE employ a two-tower architecture~\cite{karpukhin2020dense} and encode items using attributes and identifiers \& attributes, respectively.
As shown in our evaluation results, T5-ID returns an average of 49.5\% of popular products in the recommendation results of all datasets, showing that it faces the popularity bias problem during recommending items. TASTE alleviates the popularity bias by reducing on average 18.75\% popular items in its recommendation results. It represents items using full texts and utilizes text matching to calibrate the popularity-oriented recommendation behavior of T5-ID. TASTE demonstrates its effectiveness in recommending more appropriate and text-relevant items by achieving higher Bleu scores and Recall scores. Besides, compared with TASTE w/o ID, TASTE achieves higher Bleu and Dist scores with the help of item ids. It shows that the item ids can serve as a kind of prompt to provide additional matching signals beyond item attributes to better model the relevance between users and items.

\subsection{Effectiveness of TASTE on Long-Tail Items and Long-Term Interactions}\label{sec:scenarios}
This experiment evaluates the effectiveness of TASTE on modeling long-tail items and long-term user-item interactions.

\begin{figure}[t]
    \centering 
    \subfigure[Yelp.] { 
    \label{fig:line:yelp} 
     \includegraphics[width=0.48\linewidth]{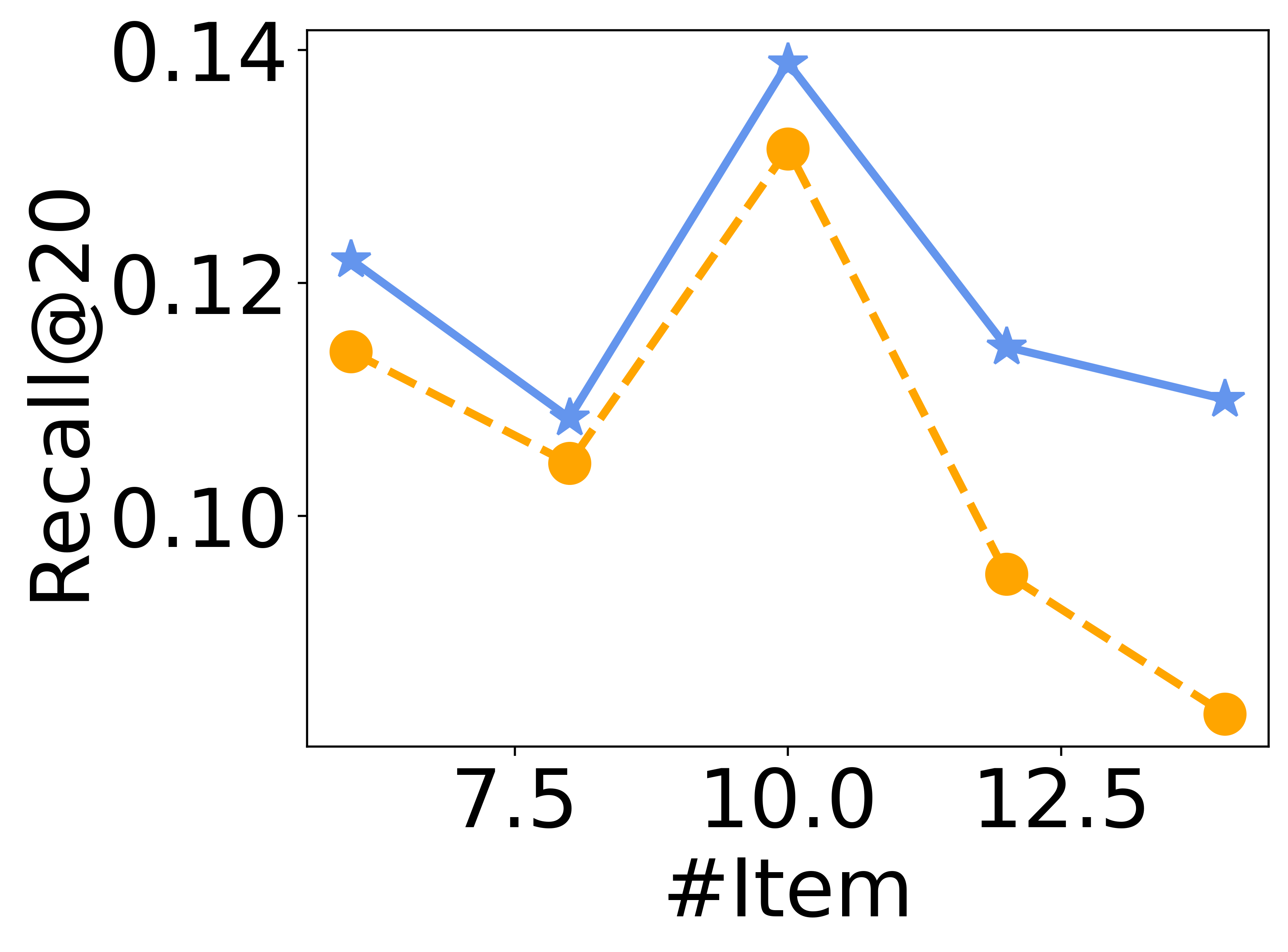}}
     \subfigure[Sports.] { 
     \label{fig:line:sports}
    \includegraphics[width=0.48\linewidth]{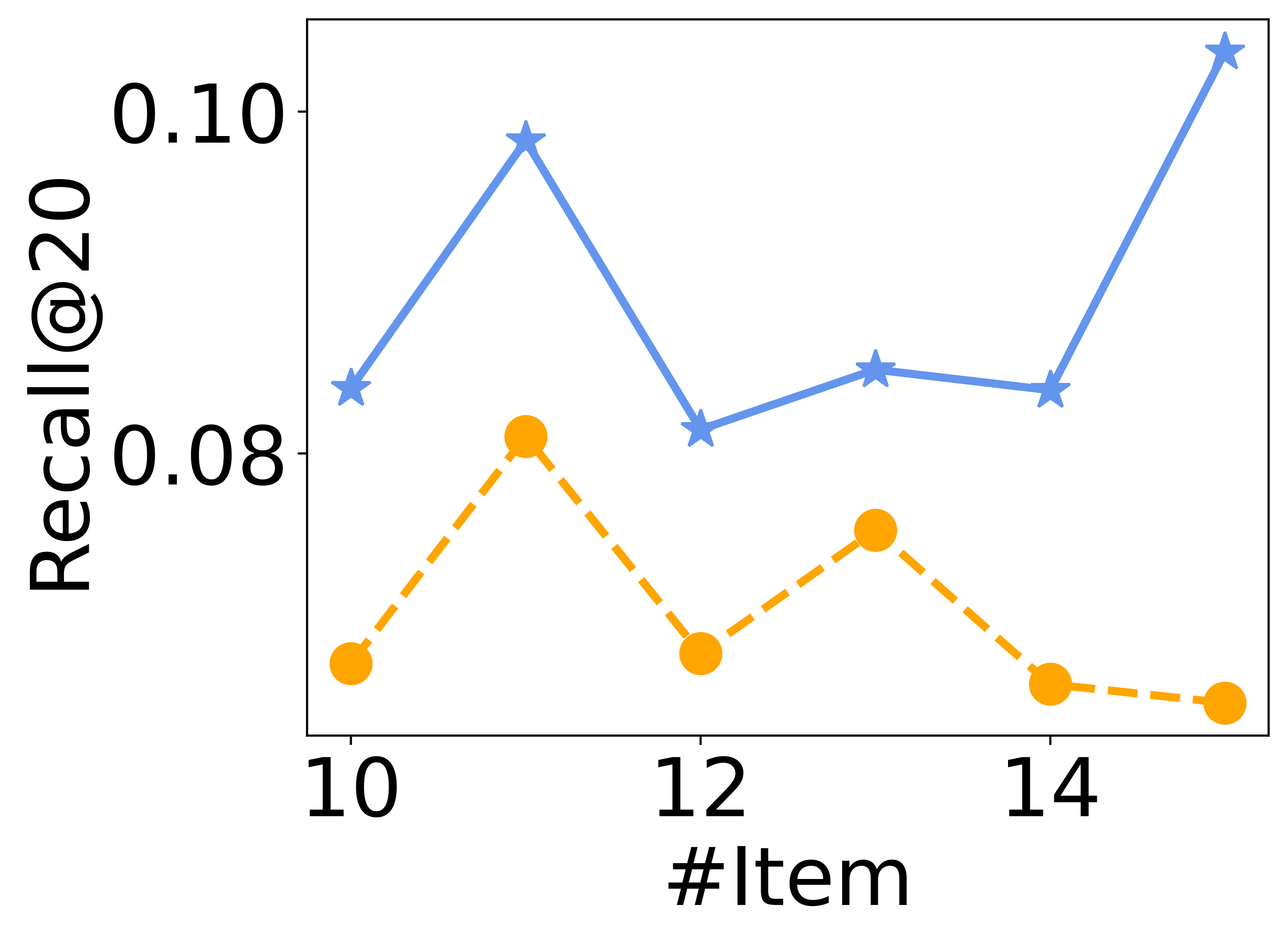}}
    \subfigure[Beauty.] { 
     \label{fig:line:beauty}
    \includegraphics[width=0.48\linewidth]{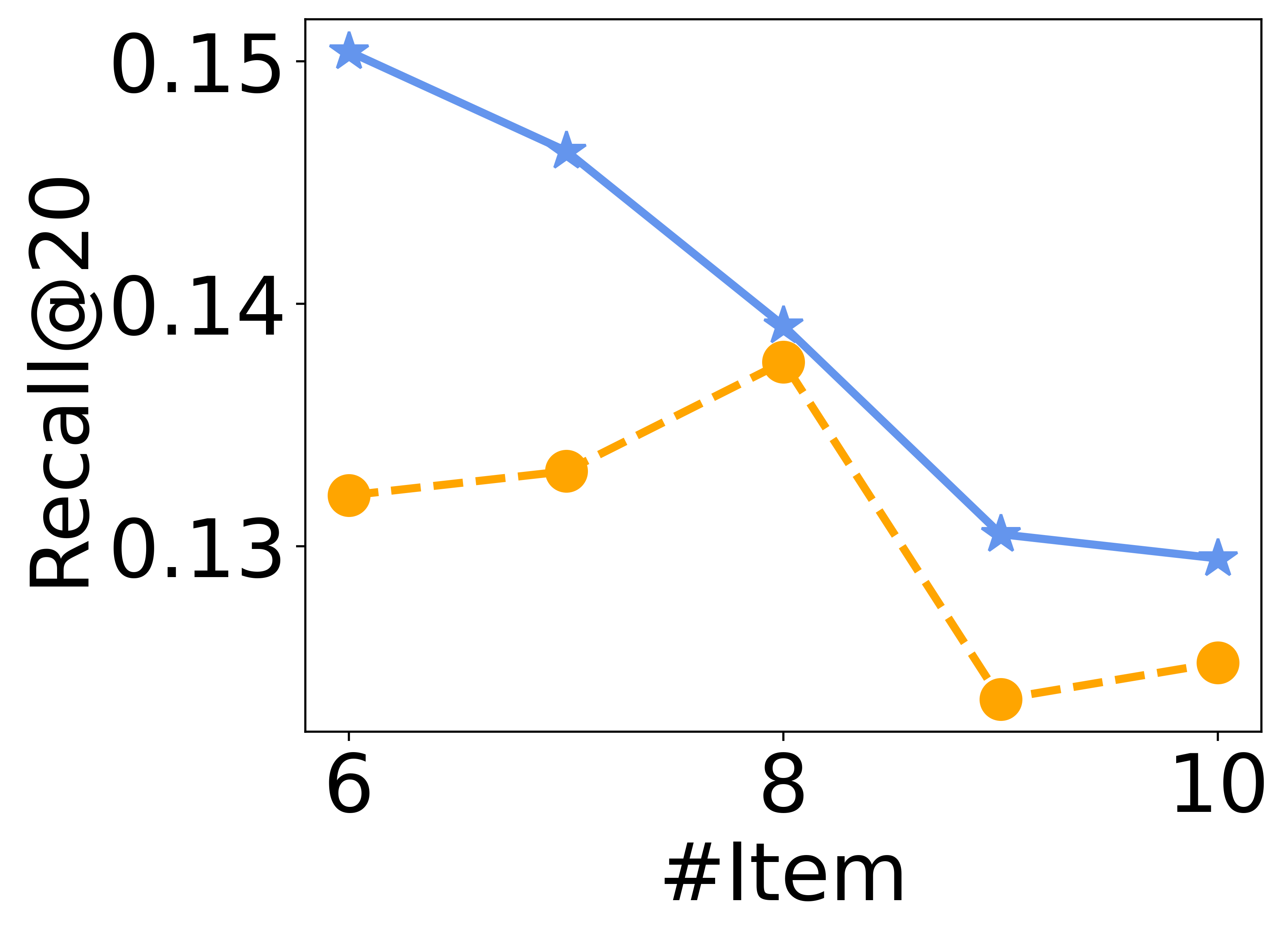}}
    \subfigure[Toys.] { 
     \label{fig:line:toys}
    \includegraphics[width=0.48\linewidth]{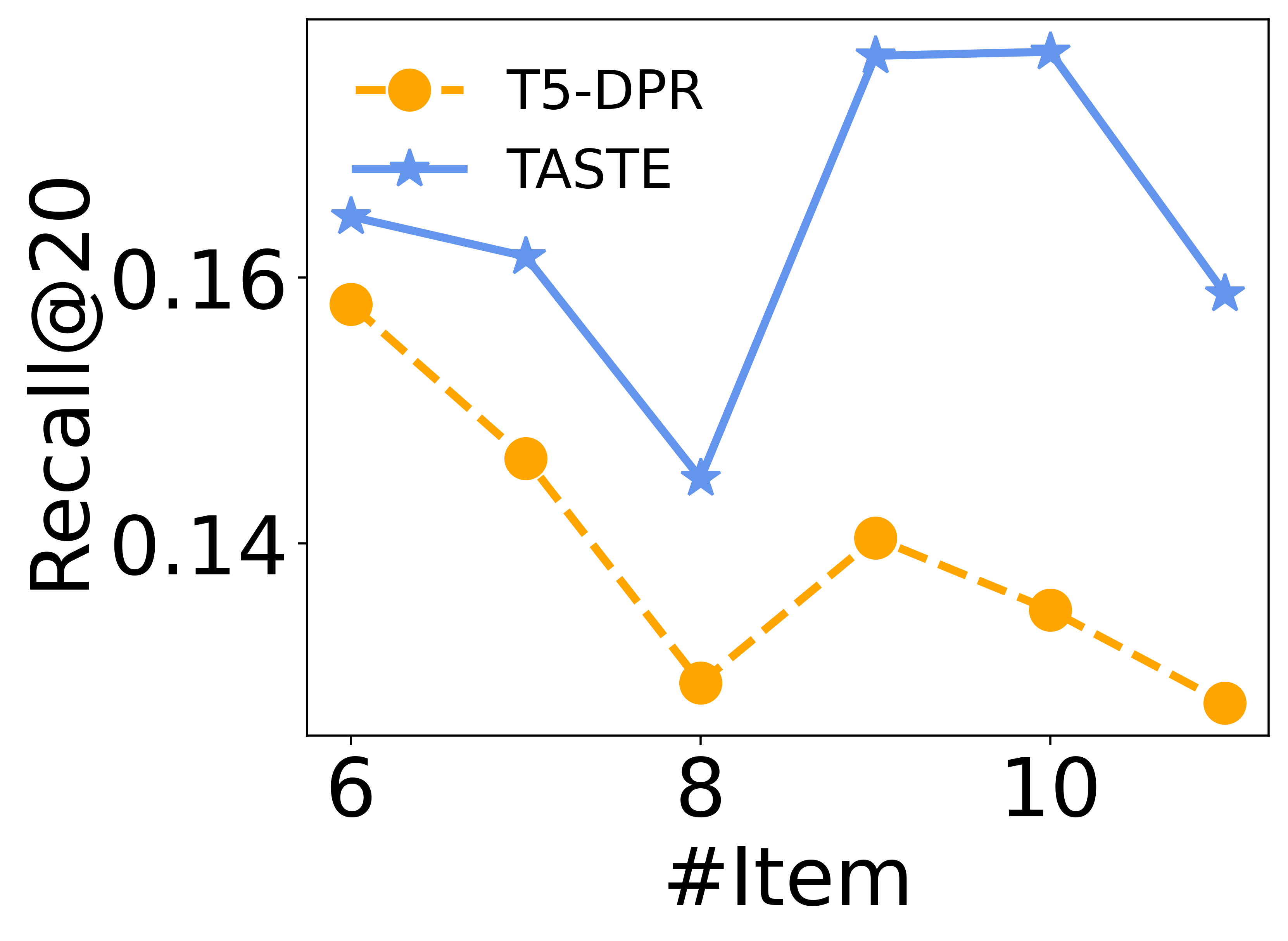}}
    \caption{Effectiveness of TASTE with Different Numbers of User-Interacted Items. Recall@20 scores are plotted along different numbers of user-interacted items.}
    \vspace{-1em}
    \label{fig:line}
\end{figure}

As shown in Figure~\ref{fig:line}, we first evaluate the recommendation performance of T5-DPR and TASTE with different numbers of interacted items. Overall, TASTE shows its advantages in modeling user-item interactions by outperforming T5-DPR with different numbers of user-interacted items. For Amazon Product datasets, TASTE achieves more improvements over T5-DPR when modeling longer user-item interactions. It thrives on our attention sparsity method and showcases the ability to accurately capture text-matching signals from longer-term user-item interactions. Different from the recommendation performance on Amazon Products, TASTE shows less effectiveness on the Yelp dataset (Figure~\ref{fig:line:yelp}) with longer user-item history. It shows that the restaurant visiting behaviors are hard to be characterized with only interacted items and should be specifically modeled for each person, such as modeling user characteristics using user identifier embeddings (Table~\ref{tab:idmodeling}).

\begin{figure}[t]
    \centering 
    \subfigure[Long-Tail Items.] { 
    \label{fig:freq:low} 
     \includegraphics[width=0.48\linewidth]{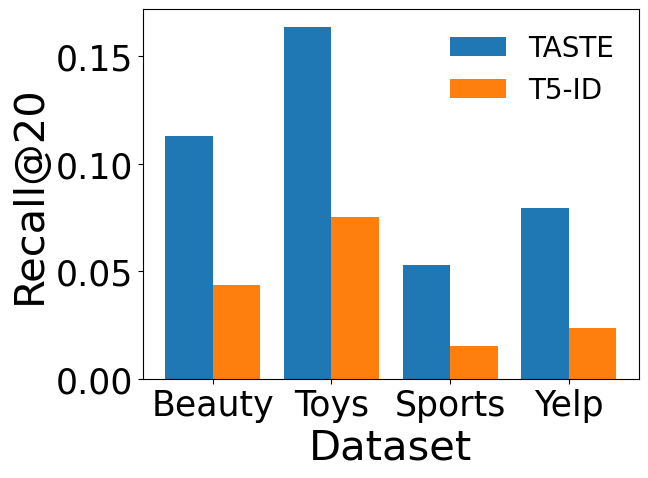}}
     \subfigure[Other Items.] { 
     \label{fig:freq:high} 
    \includegraphics[width=0.48\linewidth]{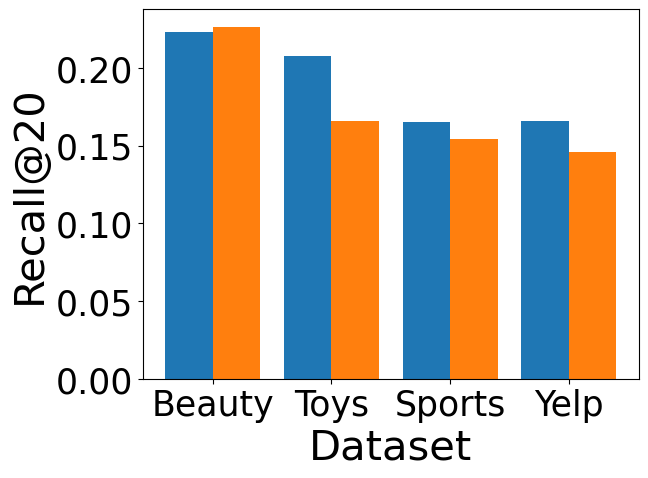}}
    \caption{Recommendation Effectiveness on Items with Different User-Interacted Frequencies. We follow~\citet{YongchunZhu2021LearningTW}, and split the testing sets into two groups according to the user-interacted frequencies of the items of test labels. The items are sorted according to user-item interaction frequencies and are grouped into long-tail items and others by setting the item number ratio to 2:8~\cite{ZhihongChen2020ESAMDD}.}
    \label{fig:freq}
\end{figure}
Then we evaluate the recommendation performance of TASTE on these items with different user-interacted frequencies in Figure~\ref{fig:freq}. The items are divided into two groups according to user-interacted frequencies, including long-tail items and others that are more frequently interacted with users. TASTE shows more significant improvements over T5-ID on these long-tail items, which illustrates its few-shot effectiveness in learning representations of items using their identifiers and attributes to alleviate the ``Cold Start''~\cite{lam2008addressing,schein2002methods} problem in recommendation systems.
%Moreover, TASTE also keeps almost consistent performance on recommending low-frequent items and higher-frequent items.
Such an encouraging phenomenon demonstrates that TASTE can broaden the advantages of pretrained language models to these long-tail items and shed some lights to build a self-contained recommendation system by directly modeling user-item relevance by text matching.

\begin{table*}[!ht]
 \caption{Case Studies. We present two cases from Beauty and Yelp and show top-5 retrieved items from TASTE, T5-ID, and TASTE w/o ID. The matched text contents are \textcolor[rgb]{0.7,0.3,0.3}{emphasized}.\label{tab:new_case_study}}
\small
\begin{tabular}{p{0.1\linewidth}|p{0.8\linewidth}} 
\hline
%\makecell[l]{\textbf{Case1}} & \makecell{Yelp}
\multicolumn{2}{l}{\textbf{Case \#1 in Beauty}}
\tabularnewline 
\hline
\makecell[l]{History} & id: 10694 title: \textcolor[rgb]{0.7,0.3,0.3}{Axe 3 in 1 Shampoo Plus Conditioner Plus Bodywash} Total Fresh, 12 Ounce, id: 4271 title: Neutrogena Ageless Essentials Continuous Hydration, Night, 1.7 Ounce, id: 3336 title: Zeno Mini Acne Clearing Device, White, id: 2778 title: \textcolor[rgb]{0.7,0.3,0.3}{Nivea For Men} Energy Hair and \textcolor[rgb]{0.7,0.3,0.3}{Body Wash}, 16.9-Ounce Bottle \textcolor[rgb]{0.7,0.3,0.3}{(Pack of 3)}, id: 3182 title: Axe Detailer \textcolor[rgb]{0.7,0.3,0.3}{Shower Tool}-Colors May Vary
\tabularnewline 
\hline
\makecell[l]{Label} & \textcolor[rgb]{0.7,0.3,0.3}{id: 10918 title: Nivea 3-in-1 Pure Impact Body Wash for Men, 16.9 Ounce (Pack of 3)}
\tabularnewline 
\hline
\makecell[l]{T5-ID}
&
\makecell[l]{\textit{1st:} id: 10844 title: \textcolor[rgb]{0.7,0.3,0.3}{Axe} Shower Gel Deep Space, 16 Ounce \\ \textit{2nd:} id: 10811 title: \textcolor[rgb]{0.7,0.3,0.3}{Axe} Shower Gel Apollo, 16 Ounce \\ \textit{3rd:} 10538 title: Dove Men+Care Sensitive + Face Lotion 1.69 FL.Oz. \\ \textit{4th:} 10701 title: Dove Men+Care Anti Dandruff Fortifying Shampoo, 12-Ounces \\ \textit{5th:} 10570 title: \textcolor[rgb]{0.7,0.3,0.3}{Axe} Energizing Face Wash, Boost, 5 Ounce}
\tabularnewline 
\hline
\makecell[l]{TASTE w/o ID}
&
\makecell[l]{\textit{1st:} id: 10694 title: \textcolor[rgb]{0.7,0.3,0.3}{Axe 3 in 1 Shampoo Plus Conditioner Plus Bodywash Total Fresh, 12 Ounce} \\ \textit{2nd:} id: 10693 title: \textcolor[rgb]{0.7,0.3,0.3}{Axe 2 in 1 Shampoo Plus Conditioner} Apollo, 12 Ounce \\ \textit{3rd:} id: 10811 title: \textcolor[rgb]{0.7,0.3,0.3}{Axe} Shower Gel Apollo, 16 Ounce \\ \textit{4th:} id: 10701 title: Dove Men+Care Anti Dandruff Fortifying \textcolor[rgb]{0.7,0.3,0.3}{Shampoo}, 12-Ounces \\ \textit{5th:} id: 10698 title: \textcolor[rgb]{0.7,0.3,0.3}{Axe} Anti-dandruff Styling Cream, 3.2 Ounce}
\tabularnewline 
\hline
\makecell[l]{TASTE}
&
\makecell[l]{\textit{1st:} \textcolor[rgb]{0.7,0.3,0.3}{id: 10918 title: Nivea 3-in-1 Pure Impact Body Wash for Men, 16.9 Ounce (Pack of 3)} \\ \textit{2nd:} id: 5293 title: Dove \textcolor[rgb]{0.7,0.3,0.3}{Men + Care Body} and Face Bar, Extra Fresh, 4 Ounce, 8 Count \\ \textit{3rd:} id: 2778 title: \textcolor[rgb]{0.7,0.3,0.3}{Nivea For Men Energy Hair and Body Wash}, 16.9-Ounce Bottle (Pack of 3) \\ \textit{4th:} id: 4205 title: \textcolor[rgb]{0.7,0.3,0.3}{Nivea For Men Active3 Body Wash} for Body, Hair  Shave \\ \textit{5th:} id: 2776 title: \textcolor[rgb]{0.7,0.3,0.3}{Nivea For Men Sensitive Body Wash} 3-in-1 Body, Hair  Face}
\tabularnewline 
\hline
\multicolumn{2}{l}{\textbf{Case \#2 in Yelp}}
\tabularnewline 
\hline
\makecell[l]{History} & id: 5908 title: DoMazing address: 6659 S Las Vegas Blvd, Ste B-101 Las Vegas NV, id: 2053 title: Village Pub \& Poker address: \textcolor[rgb]{0.7,0.3,0.3}{7575 S Rainbow Blvd Las Vegas NV}, id: 15510 title: \textcolor[rgb]{0.7,0.3,0.3}{Starbucks} address: 7855 Blue Diamond Rd, 101 Las Vegas NV, id: 8414 title: Bowlology address: 4680 Maryland Pkwy, Ste 102 Las Vegas NV, id: 1477 title: \textcolor[rgb]{0.7,0.3,0.3}{IKEA - Cafe} address: 6500 Ikea Way Las Vegas NV, $...$ , id: 16133 title: Meet Fresh address: 3930 Spring Mountain Rd Las Vegas NV, id: 12190 title: Yanni's Greek Grill address: 9620 S Las Vegas Blvd, Ste E-7 Las Vegas NV
\tabularnewline 
\hline
Label & \textcolor[rgb]{0.7,0.3,0.3}{id: 15108 title: Black Rock Coffee Bar address: 7565 S Rainbow Blvd Las Vegas NV}
\tabularnewline 
\hline
T5-ID
&
\makecell[l]{\textit{1st:} id: 195 title: Outback Steakhouse address: 7380 Las Vegas Blvd S Las Vegas NV \\ \textit{2nd:} id: 17172 title: Myungrang Hot Dog address: 4284 Spring Mountain Rd, D101 Las Vegas NV \\ \textit{3rd:} id: 6201 title: Dutch Bros \textcolor[rgb]{0.7,0.3,0.3}{Coffee} address: 4585 Blue Diamond Rd Las Vegas NV \\ \textit{4th:} id: 12468 title: Brother's Pizza address: 7575 S Rainbow Blvd, Ste 104 Las Vegas NV \\ \textit{5th:} id: 13175 title: Graffiti Bao address: 7355 S Buffalo Dr, Ste 1 Las Vegas NV}
\tabularnewline 
\hline
TASTE w/o ID
&
\makecell[l]{\textit{1st:} \textcolor[rgb]{0.7,0.3,0.3}{id: 15108 title: Black Rock Coffee Bar address: 7565 S Rainbow Blvd Las Vegas NV}  \\ \textit{2nd:} id: 15011 title: La Luna Tea and Dessert Bar address: Ste 8000F Las Vegas NV \\ \textit{3rd:} id: 2836 title: \textcolor[rgb]{0.7,0.3,0.3}{Mazzoa Donuts} address: 5180 Blue Diamond Rd, Ste 110 Las Vegas NV \\ \textit{4th:} id: 373 title: Blaze Fast-Fire'd Pizza address: 8229 Dean Martin Dr, Ste 115 Las Vegas NV \\ \textit{5th:} id: 8355 title: \textcolor[rgb]{0.7,0.3,0.3}{Starbucks} address: 8263 Dean Martin Drive Las Vegas NV}
\tabularnewline 
\hline 
TASTE 
&
\makecell[l]{\textit{1st:} id: 15108 title: \textcolor[rgb]{0.7,0.3,0.3}{Black Rock Coffee Bar address: 7565 S Rainbow Blvd Las Vegas NV}  \\ \textit{2nd:} id: 5908 title: \textcolor[rgb]{0.7,0.3,0.3}{DoMazing} address: 6659 S Las Vegas Blvd, Ste B-101 Las Vegas NV \\ \textit{3rd:} id: 4838 title: \textcolor[rgb]{0.7,0.3,0.3}{Babystacks Cafe} address: 8090 Blue Diamond Rd, Ste 170 Las Vegas NV \\ \textit{4th:} id: 17094 title: \textcolor[rgb]{0.7,0.3,0.3}{Black Rock Coffee Bar} address: 4835 Blue Diamond Rd Las Vegas NV \\ \textit{5th:} id: 2836 title: \textcolor[rgb]{0.7,0.3,0.3}{Mazzoa Donuts} address: 5180 Blue Diamond Rd, Ste 110 Las Vegas NV}
\tabularnewline 
\hline 
\end{tabular}

\end{table*}

\subsection{Case Studies}\label{app:case}
In Table~\ref{tab:new_case_study}, we present two cases from Amazon Products and Yelp to study the recommendation effectiveness of TASTE. The top-5 predicted items of T5-ID, TASTE w/o ID, and TASTE are shown. 

As shown in the first case, the user usually buys some ``Shampoo'' and ``Body Wash'' products for men. During predicting the next item, T5-ID, TASTE w/o ID, and TASTE show distinct recommendation behaviors. T5-ID prefers to predict items that are more related with the last user-interacted item and returns a series of products of the same brand ``Axe''. Compared with T5-ID, TASTE w/o ID has the ability to recommend more text-relevant items by matching the text representations of users and items. It makes TASTE w/o ID prefer to predict the historically bought items, leading to conventional recommendation results. After adding item ids as prompts, TASTE has the ability to better model item dependencies by recognizing that the user has bought a ``Shower Tool'' as the last step and the next item should be related to ``Body Wash'' instead of ``Shampoo''. Besides, compared with T5-ID, TASTE can better pick up the text clues to better model the relevance between items and shopping behavior, such as ``men'', ``3 in 1'', and ``Pack of 3''.

In the second case, we can find that the user usually visits the coffee bar and dessert bar, such as ``Starbucks '', ``DoMazing '' and ``Meet Fresh''. Unfortunately, the T5-ID model seems to fail to learn such user characteristics by returning the restaurants of ``Steakhouse'', ``Hot Dog'' and ``Pizza''. Thanks to our text matching based recommendation modeling method, TASTE can accurately capture such user preferences and assign higher recommendation ranks to coffee bars and dessert shops. It further confirms that the item dependency and user-item relevance can be fully modeled by TASTE, which generalizes items in the text space and employs pretrained language models to model text relevance between items and users.

\section{Conclusion}
In this paper, we propose a \textbf{T}ext m\textbf{A}tching based \textbf{S}equen\textbf{T}ial r\textbf{E}commendation (TASTE) model, which represents users and items with text utterances and leans text matching signals to model the relevance between them. TASTE achieves the state-of-the-art on widely used sequential recommendation datasets. It outperforms previous item id base methods by alleviating the popularity bias and better representing the long-tail items using their ids and attributes. Notably, TASTE has the ability to better model user behaviors according to long-term user-item interactions and return more text-relevant and diverse items to satisfy user needs.

\section*{Acknowledgments}
This work is supported by the Natural Science Foundation of China under Grant No. 62206042, No. 62137001, No. 61991404, and No. 62272093, the Fundamental Research Funds for the Central Universities under Grant No. N2216013, China Postdoctoral Science Foundation under Grant No. 2022M710022, and National Science and Technology Major Project (J2019-IV-0002-0069).

\bibliographystyle{ACM-Reference-Format}
\bibliography{citation}

\end{document}